\documentclass[useAMS,usenatbib]{mn2e}
\usepackage{epsfig}
\usepackage{graphicx}
\usepackage{amssymb}
\usepackage{color}
\usepackage{subfigure}
\usepackage{placeins}
\usepackage[fleqn]{amsmath}

\title[Constraining the redshifts of TeV BL Lac objects]
  {Constraining the redshifts of TeV BL Lac objects}

\author[Qin et al.]
  {Qin, Longhua$^{1,2,3,4}$\thanks{E-mail: qlh@ynao.ac.cn},
    Wang, Jiancheng$^{1,2,4}$, Yan,Dahai$^{1,2,4}$, Yang, Chuyuan$^{1,2,4}$,  \\
     \newauthor Yuan, Zunli$^{1,2,4}$, and Zhou, Ming$^{1,2,4}$\\
     $^1$Yunnan Observatory, Chinese Academy of Sciences, Kunming,
Yunnan Province 650011, PR China\\
    $^2$Key Laboratory for the Structure and Evolution of Celestial Objects,
Chinese Academy of Sciences,  Kunming, PR China\\
   $^3$University of Chinese Academy of Sciences,Beijing, PR China \\
   $^4$Center for Astronomical Mega-Science, Chinese Academy of Sciences, 20A Datun Road, Chaoyang District, Beijing, 100012, China}

\begin{document}
\pagerange{\pageref{firstpage}--\pageref{lastpage}} \pubyear{2013}

\maketitle

\label{firstpage}

\begin{abstract}

We present a model-dependent method to estimate the redshifts of three TeV BL Lac objects (BL Lacs) through fitting their (quasi-) simultaneous multi-waveband spectral energy distributions (SEDs) by one-zone leptonic synchrotron self-Compton (SSC) model. Considering the impact of electron energy distributions (EEDs) on the results, we use three types of EEDs, such as the power-law EED with exponential cut-off (PLC), the log-parabola (PLLP) and the broken power-law (BPL) EEDs, to fit the SEDs. We also use a parameter $\alpha$ to describe the uncertainties of the extragalactic background light (EBL) models, as in \citep{abd2010b}. We then use Markov Chain Monte Carlo (MCMC) method to explore multi-dimensional parameter space and obtain the uncertainties of the model parameters based on the observational data. We apply our method to obtain the redshifts of three TeV BL Lac objects in the marginalized 68\% confidence, and find that the PLC EED does not fit the SEDs. For 3C66A, the redshift is 0.14 - 0.31 and 0.16 - 0.32 in the BPL and PLLP EEDs; for PKS1424+240, the redshift is 0.55 - 0.68 and 0.55 - 0.67 in the BPL and PLLP EEDs; for PG1553+113, the redshift is 0.22 - 0.48 and 0.22 - 0.39 in the BPL and PLLP EEDs. We also estimate the redshift of PKS1424+240 in the high stage to be 0.46 - 0.67 in the PLLP EED, roughly consistent with that in the low stage.
\end{abstract}


\section{INTRODUCTION}
BL Lac object is a subclass of radio-loud active galactic
nuclei (AGN), with weak or absent optical emission lines (EW$\leq 5 \mbox{\AA}$) \citep{urr99}. The spectral energy distributions (SEDs) are mainly dominated by non-thermal components originated from a relativistic jet aligned with our line of sight and
characterized by two bumps. The first bump (and the peak is $\nu_{\rm {pk}}$), which is located at the low-energy bands from radio through UV or X-rays, is generally explained by synchrotron emission from relativistic electrons. The second hump, which is located at the high-energy bands, from GeV to the $\gamma$-ray band, in the leptonic model, is produced by the inverse Compton (IC) scattering of the relativistic electrons \citep{der1993,bott2010}, and in the hadronic model, is explained by the emission of photons or secondary particles produced in proton-photon interactions \citep{der2012,bott2013}.

For BL Lacs without line emission spectrum, its redshift is hard to be obtained directly. VHE ($\geq$ 100 GeV) photons emitted by BL Lacs are effectively absorbed by interacting with the EBL. Motivated by this mechanism, we propose a method that use a reliable EBL model to estimate the redshift of TeV BL Lac object. In this method, the redshift is a parameter and the SEDs from optical to GeV band are fitted with one-zone SSC model, then the model SED is extrapolated into the TeV band. This extrapolation is modified by the EBL absorption which only depends on the redshift if the cosmological parameters are given. Comparing the absorbed spectrum with the observed one, we can estimate the redshift $z$. In the fitting procedure, the Markov Chain Monte Carlo (MCMC) method \citep{yan2013,ino2016} is used to explore the parameter space. Based on this method, the quasi-simultaneous multi-waveband SEDs of three BL Lacs, observed by Ultraviolet and Optical telescope (OVIO), KVA 60 cm telescope, $Swift$ X-Ray telescope, $Fermi$/LAT, MAGIC telescopes, H.E.S.S and VERITAS are fitted.

It is noted that our method relies on the emission model of intrinsic spectrum. The one-zone synchrotron self-Compton (SSC) model is used to produce the SEDs of BL Lacs \citep{Ghi2010b,zhang2012}. However, in the SSC model, the EED is important to determine the SEDs \citep{yan2013,zhou2014,zhu2016}, and indicates the acceleration and cooling processes of electrons in the jet. In this paper, three types of EEDs, such as the power-law EED with exponential cut-off (PLC), the log-parabola (PLLP) and the broken power-law (BPL) EEDs are adopted, and their physical origins are reviewed in Sec. 2. We show that the EEDs might change the shape of SEDs and then lead to the different redshift. In addition, the EBL models are linked with the absorption of TeV photons \citep{fra2008,fin2010,dwe2013,ste2016}, we use a normalization factor $\alpha$ to scale the uncertainties of the EBL models, as in \cite{abd2010b}. The observed VHE flux in the energy $E_{\rm{\gamma}}$ is given by $f_{\rm{obs}}(E_{\rm{\gamma}})=f_{\rm{int}}(E_{\rm{\gamma}})\times e^{-\alpha \rm{\tau}{(E_{\rm{\gamma}},z)}}$,
where $f_{\rm{obs}}$ and $f_{\rm{int}}$ are the observed and intrinsic flux respectively, and $\rm{\tau}{(E_{\rm{\gamma}},z)}$ is the optical depth of $E_\gamma$ photon which depends on the choice of the EBL template. In the paper, we take $\alpha$ as a free parameter which has been analyzed by many authors \citep{abd2010b,hess2013, abe2015} and use the EBL template given by \cite{fin2010}.

Throughout this work, we take the cosmological parameters, $H_0$ = 70 km s$^{-1}~$Mpc$^{-1}$, $\Omega_{\rm M}$=0.3, and $\Omega_{\rm {\Lambda}}$=0.7 to calculate the luminosity distance.

\section{METHOD}
For a TeV BL Lac object at redshift $z$, the optical depth of
$E_\gamma$ photon caused by the EBL is given by
\begin{equation}
\rm{\tau}{(E_{\rm{\gamma}},z)}  = c \pi r_e^2 (\frac{m^2 c^4}{E_{\gamma}})^2
\int_0^{z} dz {dt \over dz }
   \int_{\frac{ m^2c^4}{E_{\rm{\gamma}}  (1+z)}}^\infty d\epsilon \cdot
 \epsilon^{-2} n_(\epsilon, z) \bar{\varphi}[s_0(\epsilon)],
\end{equation}
where $n (\epsilon, z)$ is the photon number density of the EBL with
energy $\epsilon$ at redshift $z$, $r_e$ is the classical electron
radius, $s_0=\epsilon E_{\rm{\gamma}}/m^2c^4$, $\bar{\varphi}[s_0(\epsilon)]$ is a function given by \cite{gou1967}, and $\frac{dt}{dz}$ is the differential time of redshift. Here we choose the EBL model offered by \cite{fin2010} to fit the SEDs of TeV BL Lacs.

We assume that the emissions of TeV BL Lacs are explained by the SSC model \citep{fin2008}, where an emitting plasma of the jet is filled with the uniform magnetic field $B$, moving with Lorentz factor $\Gamma$ at a small angle ($\theta$) to the line of sight. The intrinsic SEDs are produced by both the synchrotron radiation and the inverse Compton (IC) emission of ultra-relativistic electrons \citep{fin2008,Ghi2010b}, and are strongly enhanced by a Doppler factor $\delta_{\rm D}$, where $\delta_{\rm D} \approx {\Gamma}$ if $\theta \approx 1 / \Gamma$.
The emitting plasma is also assumed to be a spherical region with the size $R'_{\rm b}$ calculated by $R'_{\rm b} \approx \delta_{\rm D} t_{\rm {v,min}} c(1+z)^{-1}$, where $ t_{\rm {v,min}}$ is the minimum variability time-scale. It is noted that prime quantities are defined in the rest frame of the black hole, while unprimed quantities are defined in the observer frame or the fluid spherical region's frame.

Three types of EEDs are used to fit the SEDs, indicating different physical origins. The PLC and PLLP EEDs are produced by acceleration processes in jet, and the BPL EED shows no acceleration in the emitting region.

The PLC EED from shock acceleration is given by \citep{kus2000}
\begin{equation}
N^{\prime}(\gamma^{\prime})=K'_{\rm e} (\frac{-\gamma^{\prime}}{\gamma^{\prime}_{\rm c}})^{
-s}\exp(\frac{-\gamma^{\prime}}{\gamma^{\prime}_{\rm c}})\ \ {\rm for}\
\ \gamma^{\prime}_{\rm
min}\leq\gamma^{\prime}\leq\gamma^{\prime}_{\rm max},
\end{equation}
where $s$ is the electron energy spectral index, $\gamma^{\prime}_{\rm c}$ is the high energy cut-off and $K'_{\rm e}$ is the normalization factor of the EED, $\gamma_{\rm min}^{\prime}$ and
$\gamma_{\rm max}^{\prime}$ are the minimum and maximum energies of electrons, respectively. However, recent studies show that the PLC EED can also be produced by stochastic acceleration if the EED approaches the equilibrium \citep{wei2010,yan2013,peng2014}.

The PLLP EED, produced by stochastic acceleration when the acceleration dominates over radiative cooling \citep{bec2006,tra2011}, is given by
\begin{equation}
 N^{\prime}(\gamma^{\prime})=K'_{\rm e}\left\{
 \begin{array}{ll}
\left(\frac{\gamma^{\prime}}{\gamma^{\prime}_{\rm c}}\right)^{-s} & \gamma^{\prime}_{\rm min}\leq \gamma^{\prime}\leq\gamma^{\prime}_{\rm c} \\
\left(\frac{\gamma^{\prime}}{\gamma^{\prime}_{\rm c}}\right)
 ^{-[s+r\log(\frac{\gamma^{\prime}}{\gamma^{\prime}_{\rm c}})]} &  \gamma^{\prime}_{\rm c}\leq\gamma^{\prime}\leq\gamma^{\prime}_{\rm
 max}\;,
 \end{array}
 \right.
\end{equation}
where $r$ is the curvature term of the EED.

If no acceleration exists in the emitting region, the cooled EED will have the BPL shape given by \citep{fin2008}
\begin{eqnarray}
N_{\rm e}^{\prime}(\gamma^{\prime})= K'_{\rm e}H(\gamma^{\prime};\gamma_{\rm min}^{\prime},\gamma_{\rm
max}^{\prime})\{{\gamma^{\prime
-p_1}\exp(-\gamma^{\prime}/\gamma_{\rm b}^{\prime})}
\nonumber \\
\times H[(p_{\rm 2}-p_{\rm 1})\gamma_{\rm
b}^{\prime}-\gamma^{\prime}]+[(p_{\rm 2}-p_{\rm 1})\gamma_{\rm
b}^{\prime}]^{p_{\rm 2}-p_{\rm 1}}\gamma^{\prime -p_{\rm 2}}
\nonumber \\
\times \exp(p_{\rm 1}-p_{\rm 2})H[\gamma^{\prime}-(p_{\rm
2}-p_{\rm 1})\gamma_{\rm b}^{\prime}]\},
\end{eqnarray}
where $H(x;x_{1},x_{2})$ is the Heaviside function:
$H(x;x_{1},x_{2})=1$ for $x_{1}\leq x\leq x_{2}$ and
$H(x;x_{1},x_{2})=0$ everywhere else, $\gamma'_{b}$ is the break Lorentz factor, and $p_{1,2}$ is the spectral index below and above $\gamma'_{\rm {b}}$.

As shown above, there are several parameters in the model, including the size of blob $R'_{\rm{b}}$, the magnetic field $B$, the redshift $z$, the normalization factor $\alpha$, the Doppler factor $\delta_{\rm {D}}$, and the electron spectrum. $\gamma'_{\rm {min}}$ is always poorly constrained by the SED modelling. To avoid overproducing the radio flux, we set $\gamma'_{\rm {min}}$ to be 40 to 200 used in the literature. Known $t_{\rm {v,min}}$, we can get the blob's size. For the sources without minimum variability, we simply set $t_{\rm v,min}$  to be one day \citep{Ghi1998,fos2008,cao2013}. The model is not sensitive to $\gamma'_{\rm {max}}$, which is set as $\gamma'_{\rm {max}}$ = $10^8$ \citep{zhang2012,kang2016}.

Upon the above restriction, we then use the MCMC method to explore multi-dimensional parameter space and obtain the uncertainties of the model parameters based on the observed data. This method is based on Bayesian statistics with a series of parameters (hereafter $\vec{\theta}$) upon the data (hereafter $\cal D$), and the likelihood function is
\begin{equation}
{-\rm {ln}\cal L}(D\mid \vec{\theta})~\propto {\sum\limits_{i=1}^{N}(\frac{f_{\rm i}-f_{\rm {obs}}}{\sigma_{\rm {obs}}})^2},
\end{equation}
where $f_{\rm{i}}$ is the model flux in different bands, $N$ is the number of data associated with the band, $f_{\rm {obs}}$ and $\sigma_{\rm {obs}}$  are the observed flux and variance respectively. In the paper, we run the MCMC method by the public code ``CosmoMC'',\footnote{http://cosmologist.info/cosmomc/} offered by \cite{lew2002} and \cite{mac2003}. In this code, we need to input the likelihood function and the initial values of the parameters, and set the number of the  Monte-Carlo samples. This algorithm ensures that the probability density functions (PDF) of model parameters can be asymptotically approached with the number density of samples. For details please refer the papers \citep{mac2003, yuan2011, yan2013}.

After calculation, two types of probability distributions can be obtained. The maximum probability distributions are exactly the same as the best-fit one obtained by minimizing the likelihood, and the marginalized probability distributions can reflect the confident levels (C.L.) of the parameters. When the parameters are well constrained, two types of distributions will have the similar shape and interval.


\section{APPLICATIONS}
We apply the method to constrain the redshifts of three BL Lacs, e.g., 3C66A, PKS1424-240 (with low and high stages) and PG1553+113, which are from TeVCat\footnote{http://tevcat.uchicago.edu}. Their (quasi-) simultaneous multi-waveband SEDs are adopted from the different literatures.

\subsection{3C66A}
3C66A is an IBL ($10^{14}$Hz $<~\nu_{\rm{pk}}~<10^{15}$Hz ) object with a debated redshift. \cite{mil1978} firstly proposed a tentative redshift of z = 0.44, but the Mg II emission was still not confirmed. \cite{pai2017} only gave a modest lower limit of $z> 0.10$. Some authors \citep{fin2008b, yan2010, fur2013} also tried to constrain the redshift and only gave the range from 0.10 to 1.67. In the paper, we use the simultaneous SEDs given by \cite{abd2011} and \cite{rey2009}, the timescale of emitting blob with 12 hours and $\gamma_{\rm {min}}=200$ given by \cite{zhang2012}.

\subsection{PKS1424+240}
PKS1424+240 is classified as a HSP ($10^{15}$Hz $<~\nu_{\rm{pk}}$). \cite{fur2013} reported the redshift of $z>0.6$ and \cite{sba2005} gave the redshift of $z> 0.67$. \cite{yan2010} also estimated an upper limit of 1.19. Combining the observed SEDs in the GeV and TeV bands, \cite{pra2011} fitted the EBL-corrected spectra with power laws to get $z=0.24\pm 0.05$. \cite{rov2015,rov2016} studied the redshift of its likely host galaxy to give $z= 0.601\pm 0.003$. In the paper, we use the simultaneous SEDs from VERITAS, $Fermi$/LAT, $Swift$ at two different states observed by \cite{arc2014}. We also use the timescale of X-ray variability with about a day \citep{acc2010} and $\gamma_{\rm {min}}=100$ \citep{zhang2012}. It is found that $\gamma_{\rm {min}} =40$ can also give the better result.

\subsection{PG1553+113}
The VHE observation indicates that the redshift of PG 1553+113 is greater than 0.30 \citep{lan2014}. \cite{abr2015} used Bayesian statistics to analyze HESS data during the nights of 2012 April 26 and 27 and got $z=0.49\pm 0.04$. In addition, the authors \citep{yan2010,dan2010,alu2015} also gave the upper limit of redshift from 0.40 to 0.78.
We use the quasi-simultaneous SEDs from \cite{abd2010}, the timescale of one day due to the lack of variability data, and $\gamma_{\rm {min}}=200$ \citep{zhang2012}.

\section{RESULTS}

We have used three types of EEDs to fit the SEDs of TeV BL Lacs. The SEDs in the cases of the maximum probability and the 1$\sigma$ error band are shown in Fig.1. The one-dimensional (1D) probability distributions and the two-dimensional (2D) confident contours of the model parameters are shown in Fig.2.  In the 1D distributions, the dotted lines show the maximum likelihood distributions, and the solid lines show the marginalized distributions. In the 2D distributions, the inner contour denotes the 68\% C.L., while the outer contour represents the 95\% C.L. The model parameters are listed in Table 1-3.

\begin{table*}
\centering
\small
\caption{The model parameters of the best-fits and the marginalized 68\%
confident intervals (CI) are listed for BPL EEDs. \label{Table:1}}
\begin{tabular}{lccc ccccc ll}
\hline \hline
Source name & $z$ & $ B~(0.1G)$ &Log$[{\gamma'_{\rm b}}]$ & $\delta_{\rm D}~(10)$& Log$[K'_{\rm e}]$ & $p_1$ & $p_2$ &$\alpha$& $\chi^2/{\rm {d.o.f}}$\\
~[1] & [2] & [3] & [4] & [5] & [6] &[7] & [8] & [9]&[10] \\
\hline
3C66A&0.16&0.95&4.21&2.04&50.62&1.22&5.04&1.02&1.22\\
(68\%~CI)&0.14~-~0.31&0.58~-~1.01&4.18~-~4.29&1.90~-~3.18&50.24~-~51.53&1.10~-~1.42&4.99~-~5.11&0.80~-~1.20&-\\
\\
PKS1424+240&0.67&0.11&4.54&8.83&54.27&2.01&4.95&0.53&4.39\\
(68\%~CI)&0.55~-~0.68&$<$0.15&4.41~-~4.59&7.43~-~8.74&53.34~-~54.58&1.78~-~2.11&4.66~-~5.29&0.50~-~0.67&-\\
\\
PKS1424+240 flare&0.34&0.04&5.29&9.82&56.43&2.59&4.10&1.27&7.23\\
(68\%~CI)&0.34~-~0.48&0.05~-~0.08&5.13~-~5.29&$>$ 8.79&55.97~-~56.48&2.49~-~2.61&4.00~-~4.23&$>$ 0.80 &-\\
\\
PG1553+113&0.48&0.36&4.21&6.34&49.30&0.84&3.97&0.41&1.23\\
(68\%~CI)&0.22~-~0.48&0.16~-~0.33&4.04~-~4.26&3.74~-~6.44&47.13~-~49.91&0.31~-~1.02&3.94~-~4.00&0.50~-~0.88&-\\

\hline
\end{tabular}
\vskip 0.4 true cm
\end{table*}

\textbf{\textit{3C66A}}. The SED fitted by the PLC EED does not cover the $Fermi$/LAT bands, and $\chi^2/{\rm {d.o.f}}$ is worse than that of other two EEDs. It is found that the SED fitted by the PLC EED within the 68\% C.L. does not cover $fermi$ band well comparing with other two EEDs. The PLC EEDs has then been ruled out in modeling the SEDs. We obtain that $\chi^2/{\rm {d.o.f}}$ are 1.22 and 1.20 for PBL and PLLP EEDs, and the parameters are well constrained except for $\alpha$ in both models. However, the flux in 20 KeV-50 KeV bands given by the PLLP EED is higher than that by the BPL EEDs, indicating that hard X-ray observation could distinguish two types of EEDs. In summary, the redshift of 3C66A is 0.16 and 0.24 by the PBL and PL models respectively, and the redshift of 68\% C.L. is 0.16~-~0.32, which is in the range predicted by \cite{pai2017}.

\begin{table*}
\centering
\small
\caption{The model parameters of the best-fits and the marginalized 68\%
confident intervals (CI) are listed for PLLP EEDs. \label{Table:2}}
\begin{tabular}{lccc ccccc ll}
\hline \hline
Source name & $z$ & $ B~(0.1G)$ & Log$[{\gamma'_{\rm c}}]$ & $\delta_{\rm D}~(10)$& Log$[K'_{\rm e}]$ & $s$ & $r$ &$\alpha$& $\chi^2/{\rm {d.o.f}}$\\
~[1] & [2] & [3] & [4] & [5] & [6] &[7] & [8] & [9]&[10] \\
\hline
3C66A&0.24&0.48&3.80&2.97&46.18&1.70&1.15&1.12&1.20\\
(68\%~CI)&0.16~-~0.32&0.32~-~0.71&3.42~-~3.85&2.21~-~3.84&46.05~-~46.83&1.10~-~1.20&1.03~-~1.20&0.80~-~1.20&-\\
\\
PKS1424+240&0.67&0.10&2.99&8.90&47.33&0.07&1.05&0.53&4.15\\
(68\%~CI)&0.55~-~0.67&$<$ 0.14&3.11~-~3.60&7.41~-~8.61&46.83~-~47.25&0.23~-~1.29&1.04~-~1.12&0.50~-~0.66&-\\
\\
PKS1424+240 flare&0.71&0.10&2.91&9.95&48.19&1.42&0.46&0.53&5.92\\
(68\%~CI)&0.46~-~0.67&0.07~-~0.13&2.77~-~3.33&$>$8.14&47.50~-~48.41&1.35~-~1.84&0.42~-~0.48&$<$~0.91&-\\
\\
PG1553+113&0.32&0.10&2.46&7.00&48.00&0.48&0.84&0.95&1.33\\
(68\%~CI)&0.22~-~0.39&0.09~-~0.18&2.34~-~3.19&$>$~5.87&47.27~-~48.00&0.32~-~1.26&0.52~-~0.57&0.65~-~1.42\\

\hline
\end{tabular}
\vskip 0.4 true cm
\end{table*}

\begin{table*}
\centering
\small
\caption{The model parameters of the best-fits and the marginalized 68\%
confident intervals (CI) are listed for PLC EEDs. \label{Table:3}}
\begin{tabular}{lccc cccc ll}
\hline \hline
Source name & $z$ & $ B~(0.1G)$ & Log$[{\gamma'_{\rm c}}]$ & $\delta_{\rm D}~(10)$& Log$[K'_{\rm e}]$ & $s$ &$\alpha$& $\chi^2/{\rm {d.o.f}}$\\
~[1] & [2] & [3] & [4] & [5] & [6] &	[7] & [8] & [9] \\
\hline
3C66A&0.45&0.10&4.82&7.21&54.86&2.18&0.92&3.29\\
(68\%~CI)&0.34~-~0.43&$<$~0.16&4.76~-~4.82&5.68~-~7.00&54.66~-~54.86&2.15~-~2.20&0.80~-~1.11&-\\
\\
PKS1424+240&0.57&0.10&4.70&8.40&55.38&2.28&0.63&4.57\\
(68\%~CI)&0.53~-~0.63&0.10~-~0.13&4.68~-~4.73&7.98~-~9.00&55.08~-~55.39&2.24~-~2.32&0.55~-~0.72&-\\
\\
PKS1424+240 flare&0.26&0.03&5.55&10.91&57.21&2.76&1.49&7.39\\
(68\%~CI)&0.25~-~0.30&$<$~0.13&4.68~-~4.73&10.21~11.00&57.00~-~57.33&2.73~-~2.78&$>$~1.21&-\\
\\
PG1553+113&0.16&0.03&5.77&7.97&57.07&2.72&1.96&2.33\\
(68\%~CI)&0.25~-~0.33&0.03~-~0.04&5.65~-~5.76&$>$~7.12&57.00~-~57.23&2.69~-~2.73&$>$~1.14&-\\

\hline
\end{tabular}
\vskip 0.4 true cm
\end{table*}

\textbf{\textit{PKS1424+240}}. In the low state, comparing the results by three types of EEDs, we find that $\chi^2/{\rm {d.o.f}}$ and the constraints on model parameters are comparable.
We can not distinguish three types of EEDs from modeling the SEDs directly (see Fig 1-2 and table 1-3), but we get a similar redshift, which is from 0.56 to 0.67. This result is in the range given by \cite{rov2016}. In the PLC case, it is found from Fig.1-2 that the SED of the best-fit is slightly lower than that of other two models in the GeV band. However, the SED of the 68\% C.L. by the PLC EED covers the $Fermi$/LAT band more than by other EEDs. So we can not rule out the PLC EED in the low state.

For the observed SED of PKS1424+240 in the flare state, we also perform the SED fits using three types of EEDs. The results are summarized in Fig.1-2 and table 1-3. It is found that three types of EEDs fail to cover the Gamma ray bands, where the extreme high $\delta_{\rm D}$ ($>80.0$) is needed. It is noted that the PLLP model slightly underestimates the fluxes at $Fermi$/LAT bands, but the PLLP model ($\chi^2/{\rm {d.o.f}}$ =5.92) is better than BPL ($\chi^2/{\rm {d.o.f}}$ =7.23) and PLC ($\chi^2/{\rm {d.o.f}}$ =7.39) models. We then obtain that the redshift of the 68\% C.L. is 0.46 - 0.67, which is roughly consistent with that in the low state.

\textbf{\textit{PG1553+113}}. We find that the PLC model of the best-fit is worse than other two models from table 1-3, where its SED can not cover $10^{23}$Hz-$10^{25}$Hz bands shown in Fig.4. Its SED of the 68\% C.L. poorly covers the fluxes at the $Fermi$/LAT bands, we then rule out the PLC model. In the BPL model, the redshift is 0.48, and the redshift of 68\% C.L. is 0.22~-~0.48, which is consistent with that obtained by \cite{abr2015}. For PLLP model, the redshift is 0.32 and the redshift of 68\% C.L. is 0.22~-~0.39.

\begin{figure*}

   \begin{tabular}{ll}
\includegraphics[height=5.5cm,width=9.0cm]{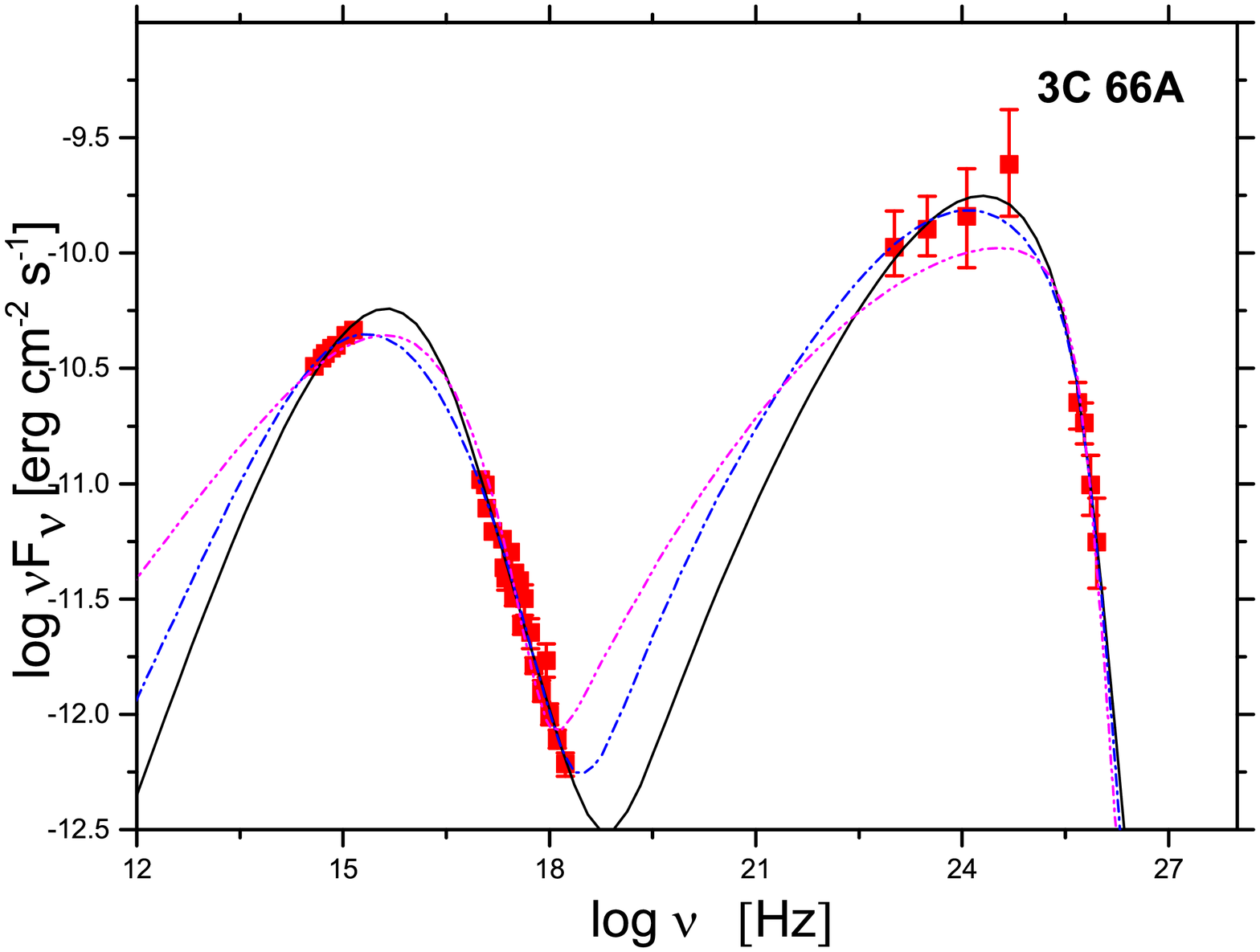}
\includegraphics[height=5.4cm,width=8.5cm]{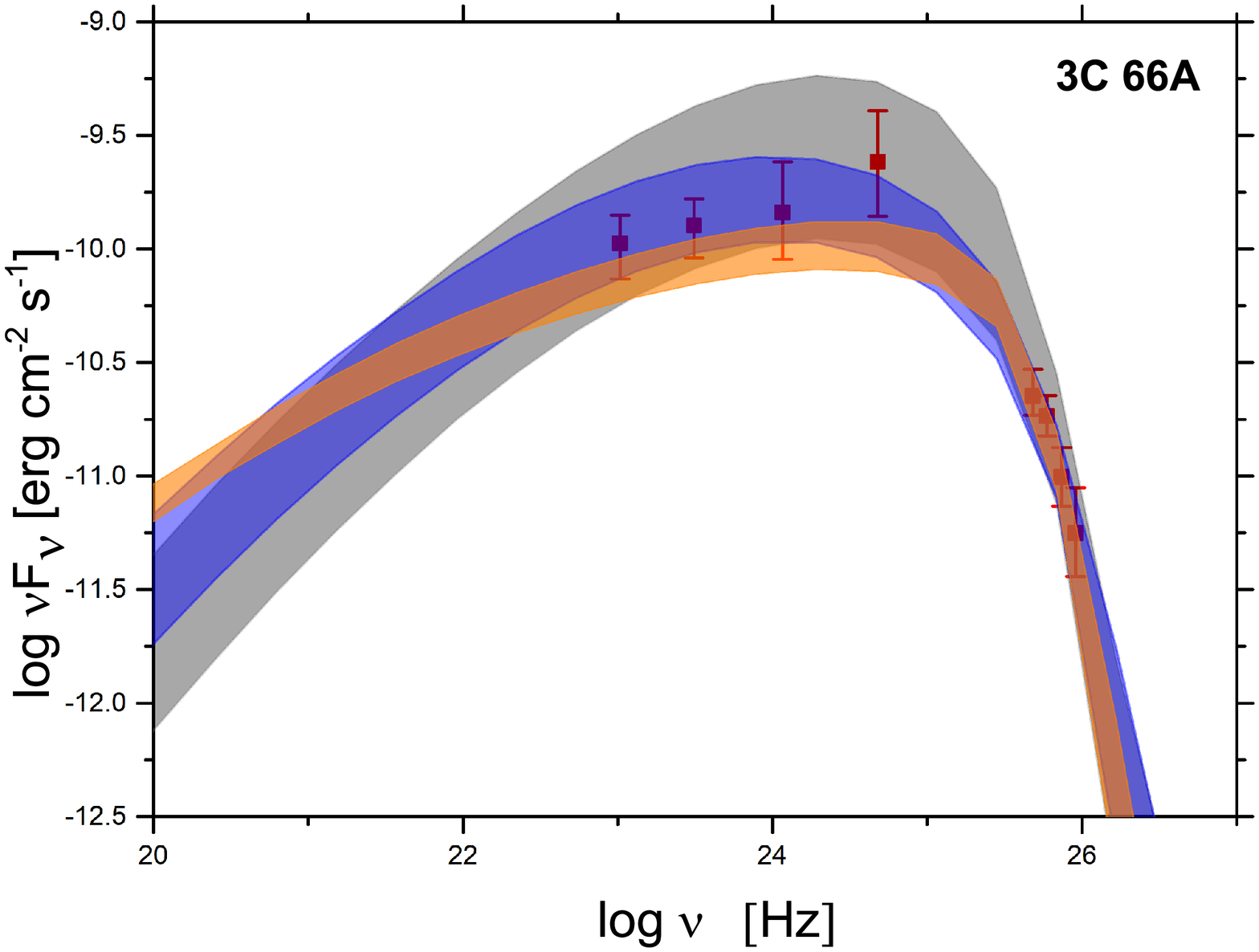}\\
\includegraphics[height=5.5cm,width=9.0cm]{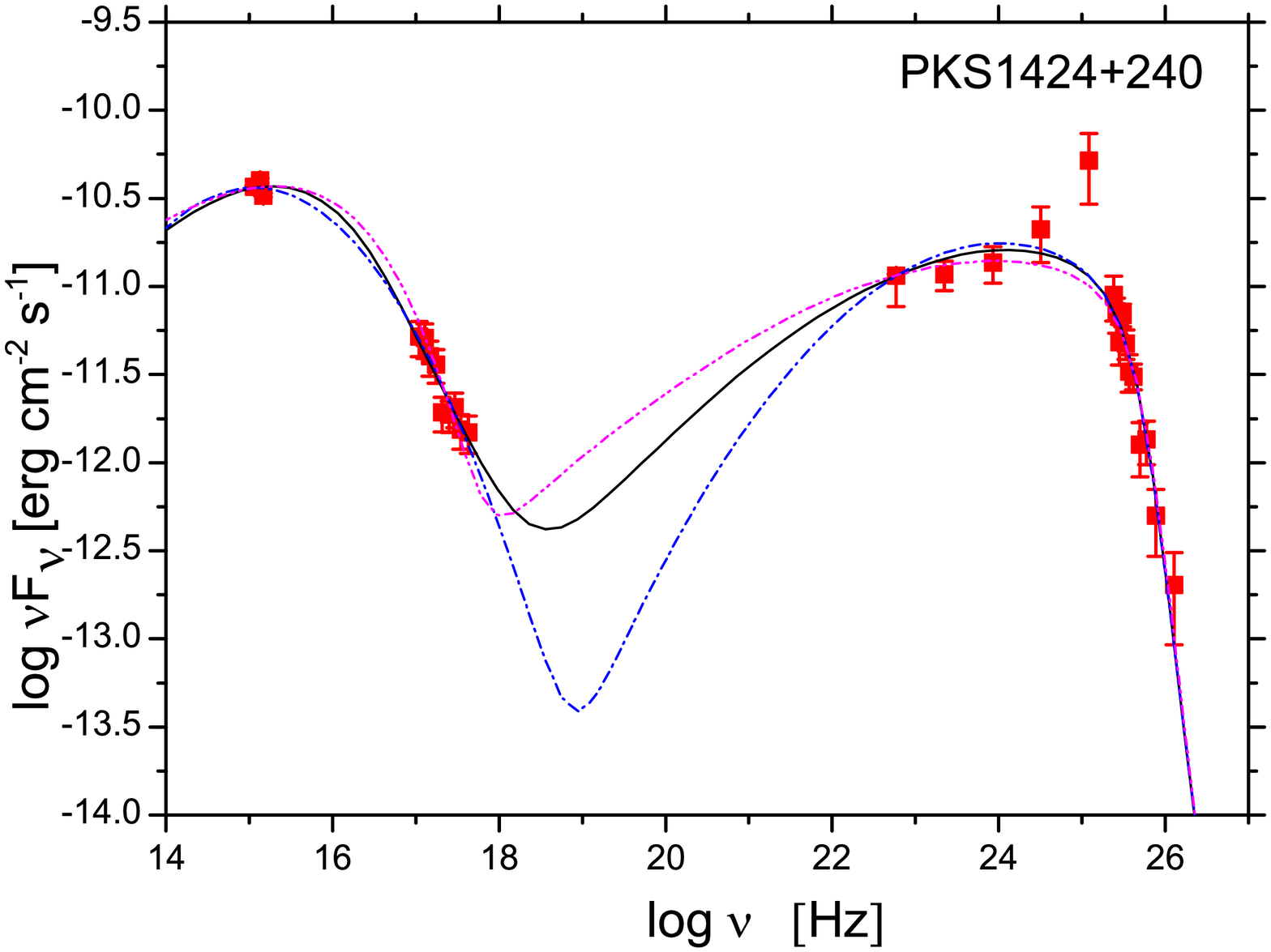}
\includegraphics[height=5.5cm,width=8.5cm]{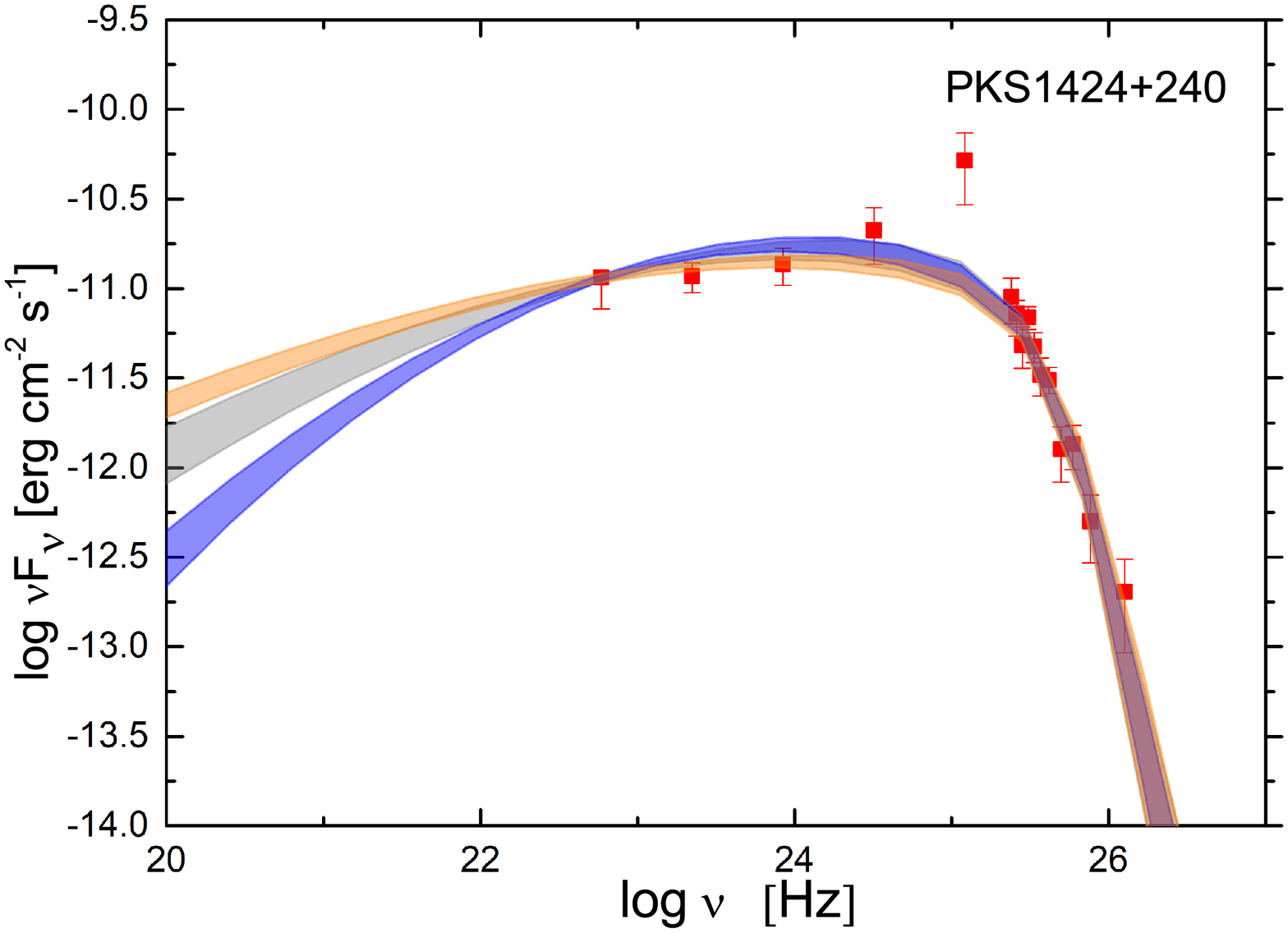}\\
\includegraphics[height=5.5cm,width=9.0cm]{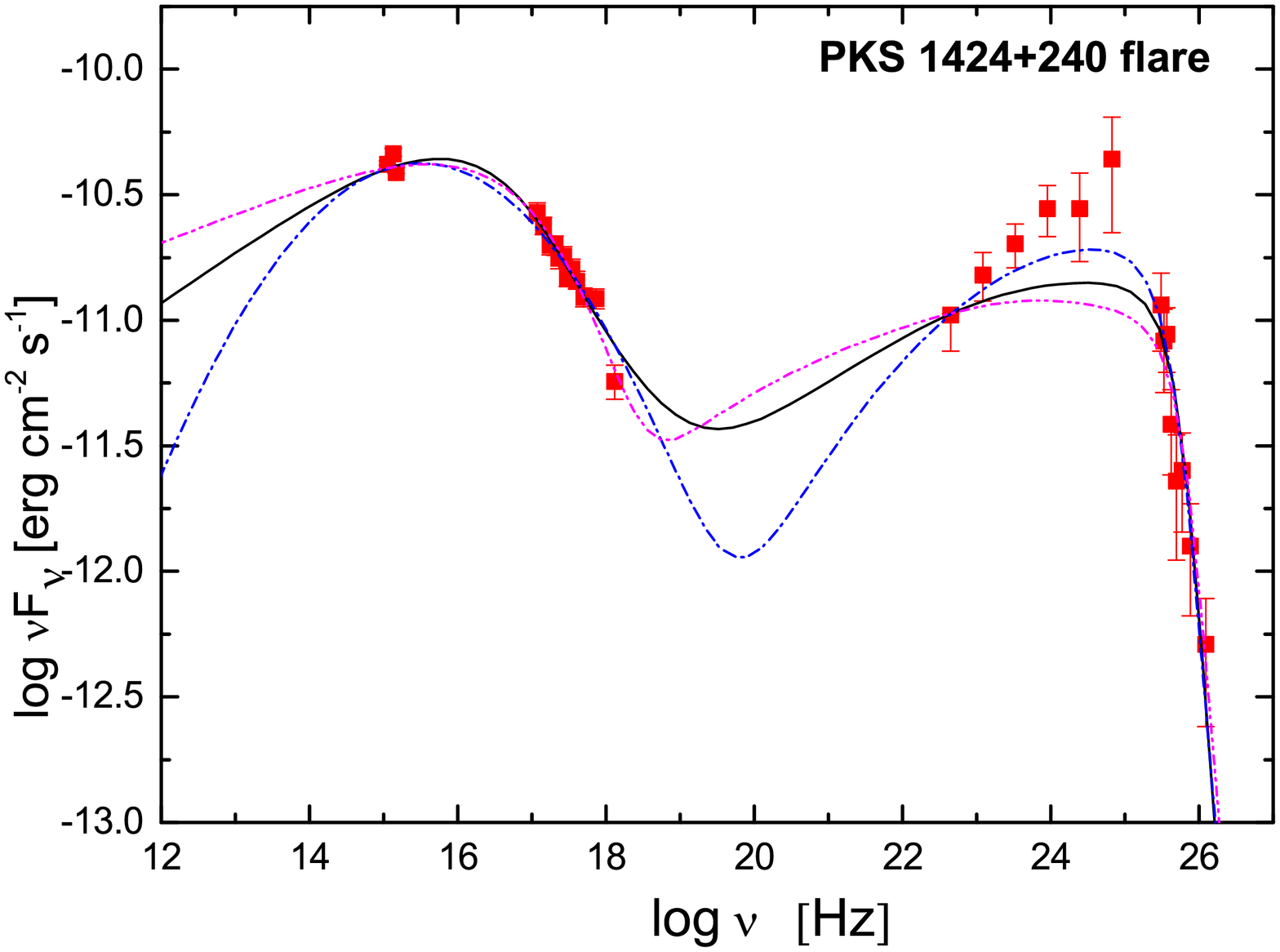}
\includegraphics[height=5.5cm,width=8.5cm]{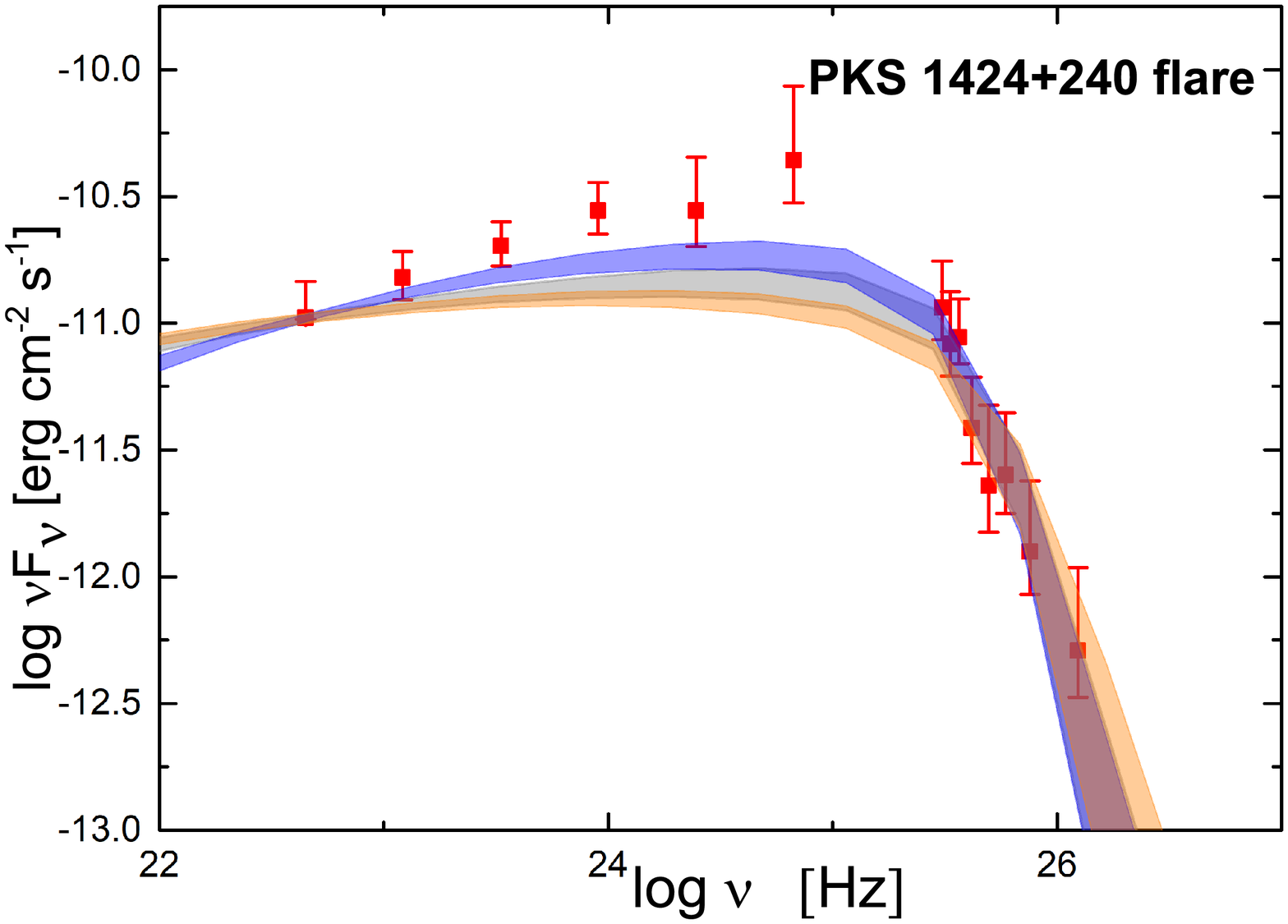}\\
\includegraphics[height=5.6cm,width=9.0cm]{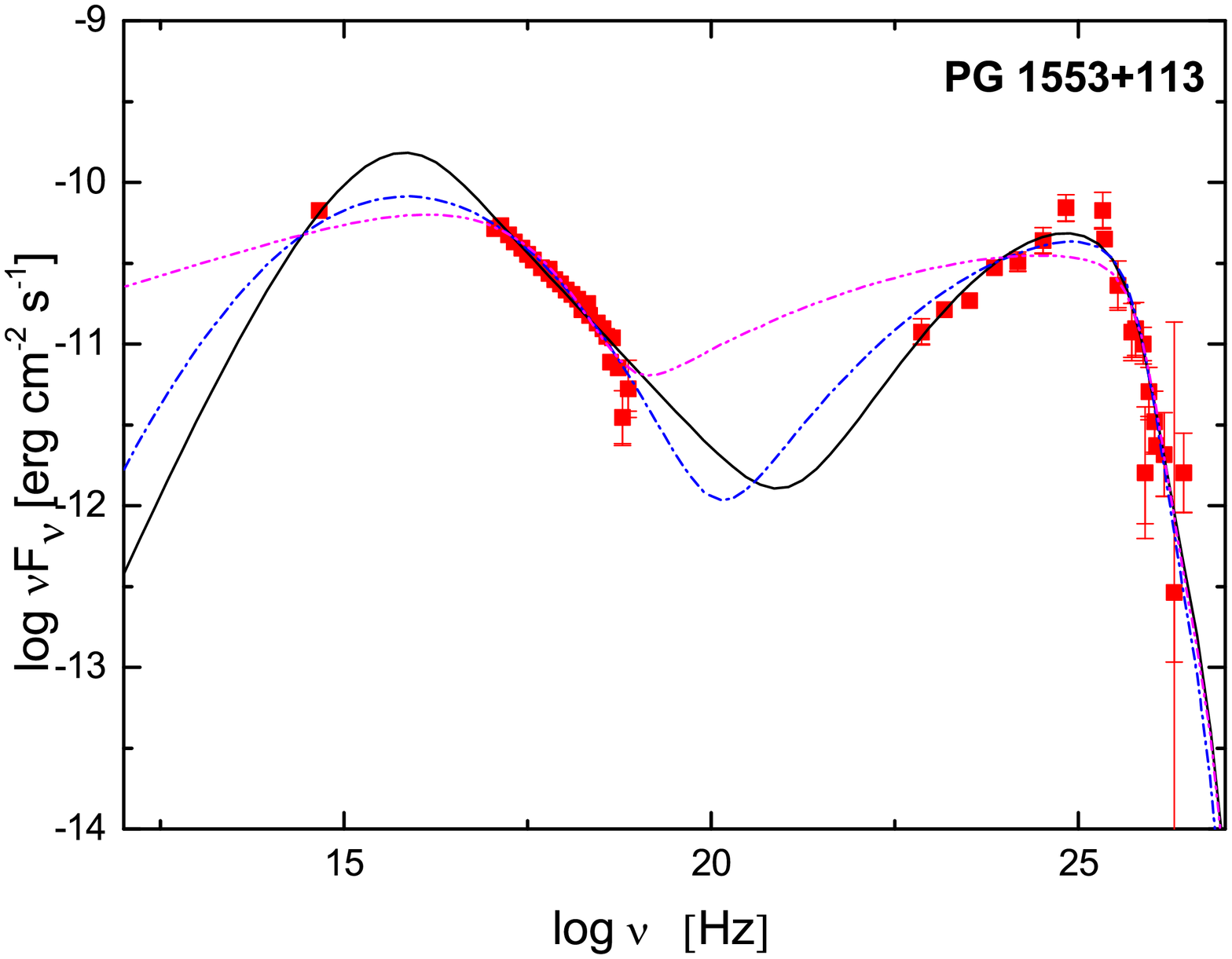}
\hfill
\includegraphics[height=5cm,width=8.5cm]{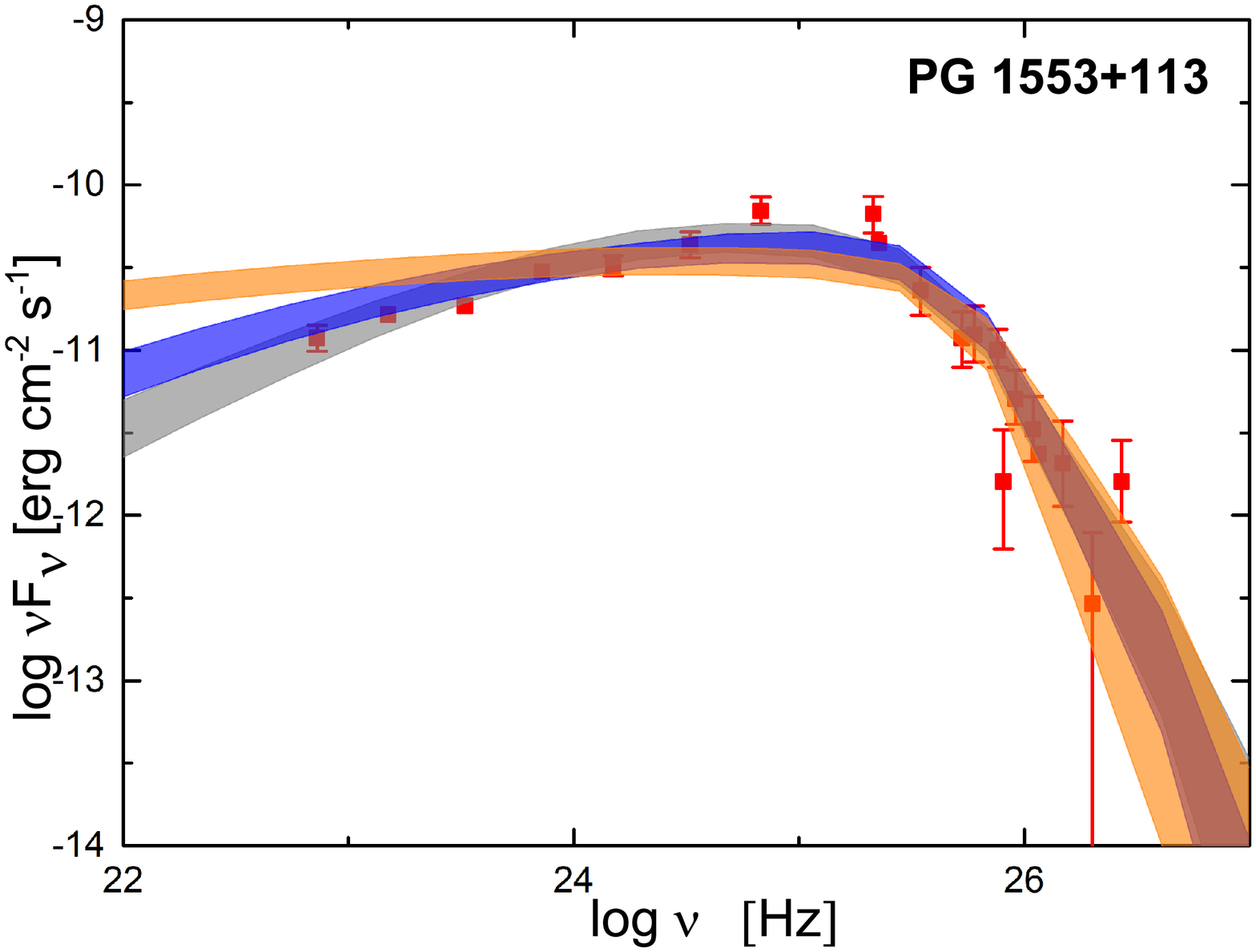}\\
    \end{tabular}
     		
\caption{Left panels: the SEDs of three objects fitted by three types of EEDs, where the sold black, blue dash dot and magenta dash dot dot line represent the SEDs given by BPL, PLLP and PLC EEDs respectively. Right panels: the SED contours of three objects in gamma-ray bands under the 1$\sigma$ error bands, where the shaded areas in gray, blue and orange color represent the SED contours given by BPL, PLLP and PLC EEDs respectively.}
\end{figure*}

\begin{figure*}
\subfigure[The distributions of the parameters in BPL EEDs]{
 \begin{minipage}[t]{8cm}
 \centering
\includegraphics[height=7cm,width=8.0cm]{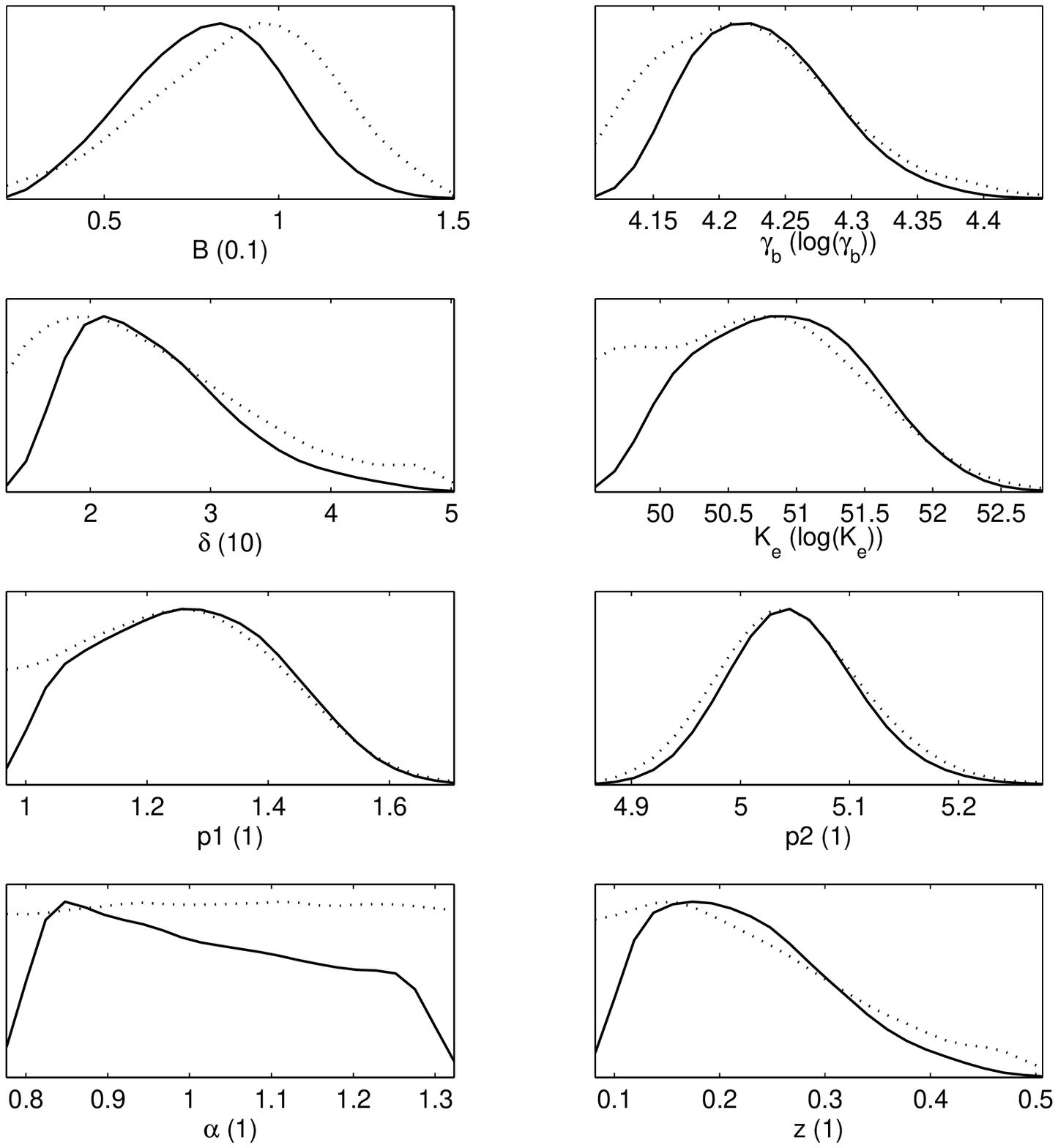}
\end{minipage}%
}
\hspace{1cm}
\subfigure[The correlations of the parameters in BPL EEDs]{
 \begin{minipage}[t]{8cm}
 \centering
\includegraphics[height=7cm,width=8.0cm]{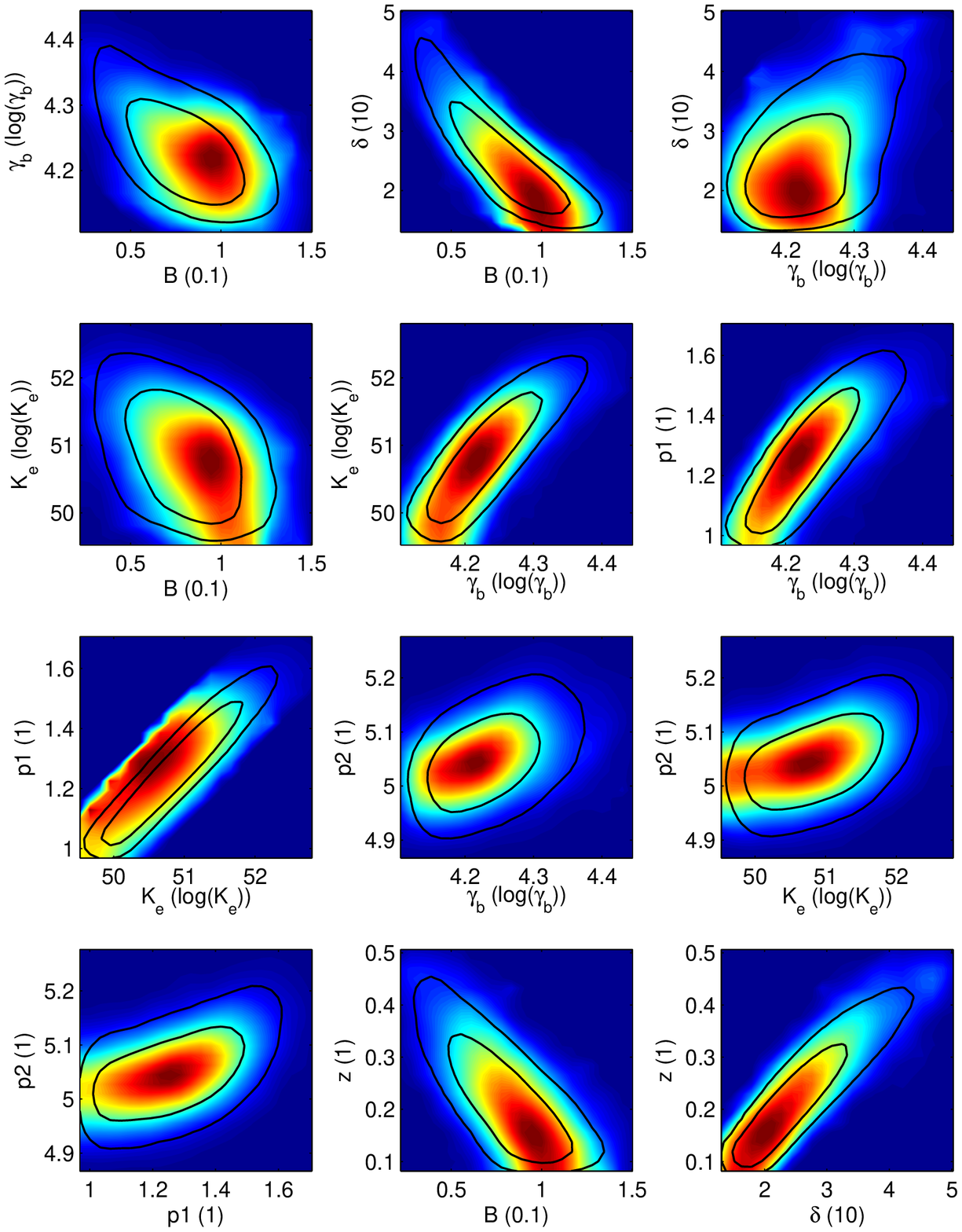}
\end{minipage}%
}
\hspace{1cm}
\subfigure[The distributions of the parameters in PLLP EEDs]{
 \begin{minipage}[t]{8cm}
 \centering
\includegraphics[height=7cm,width=8cm]{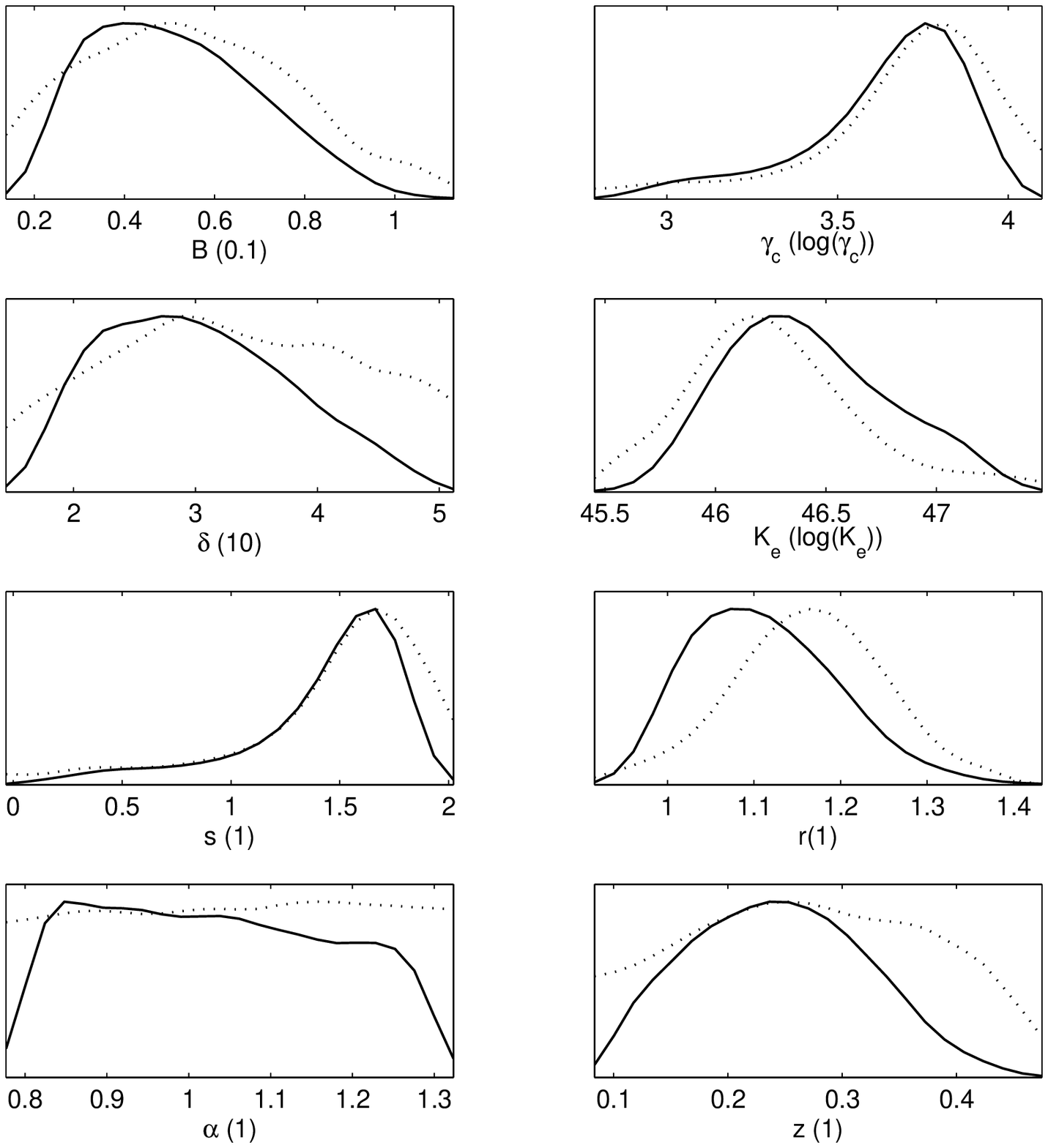}
\end{minipage}%
}
\hspace{1cm}
\subfigure[The correlations of the parameters in PLLP EEDs]{
 \begin{minipage}[t]{8cm}
 \centering
\includegraphics[height=7cm,width=8cm]{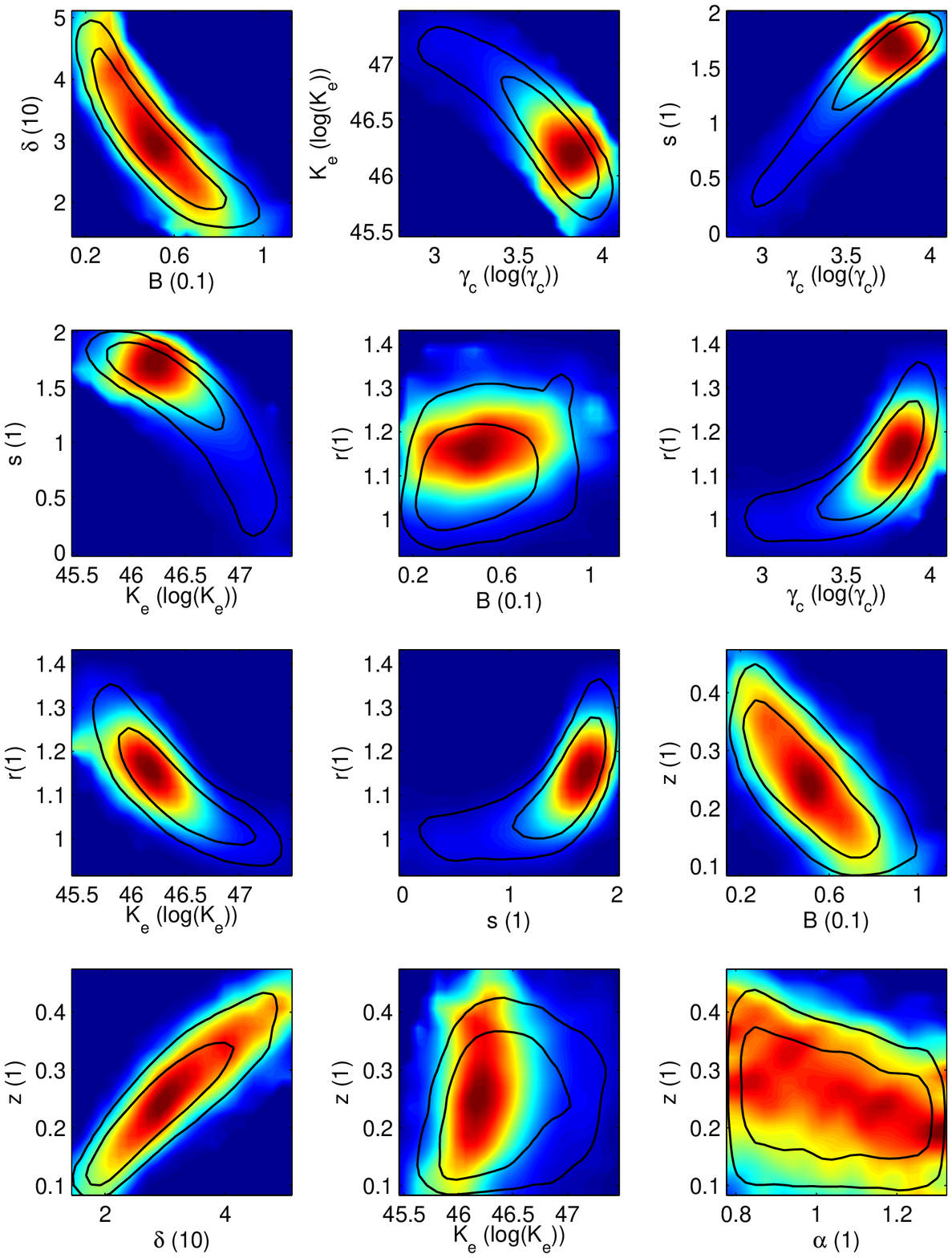}
\end{minipage}%
}
\hspace{1cm}
\subfigure[The distributions of the parameters in PLC EEDs]{
 \begin{minipage}[t]{8cm}
 \centering
\includegraphics[height=7cm,width=8cm]{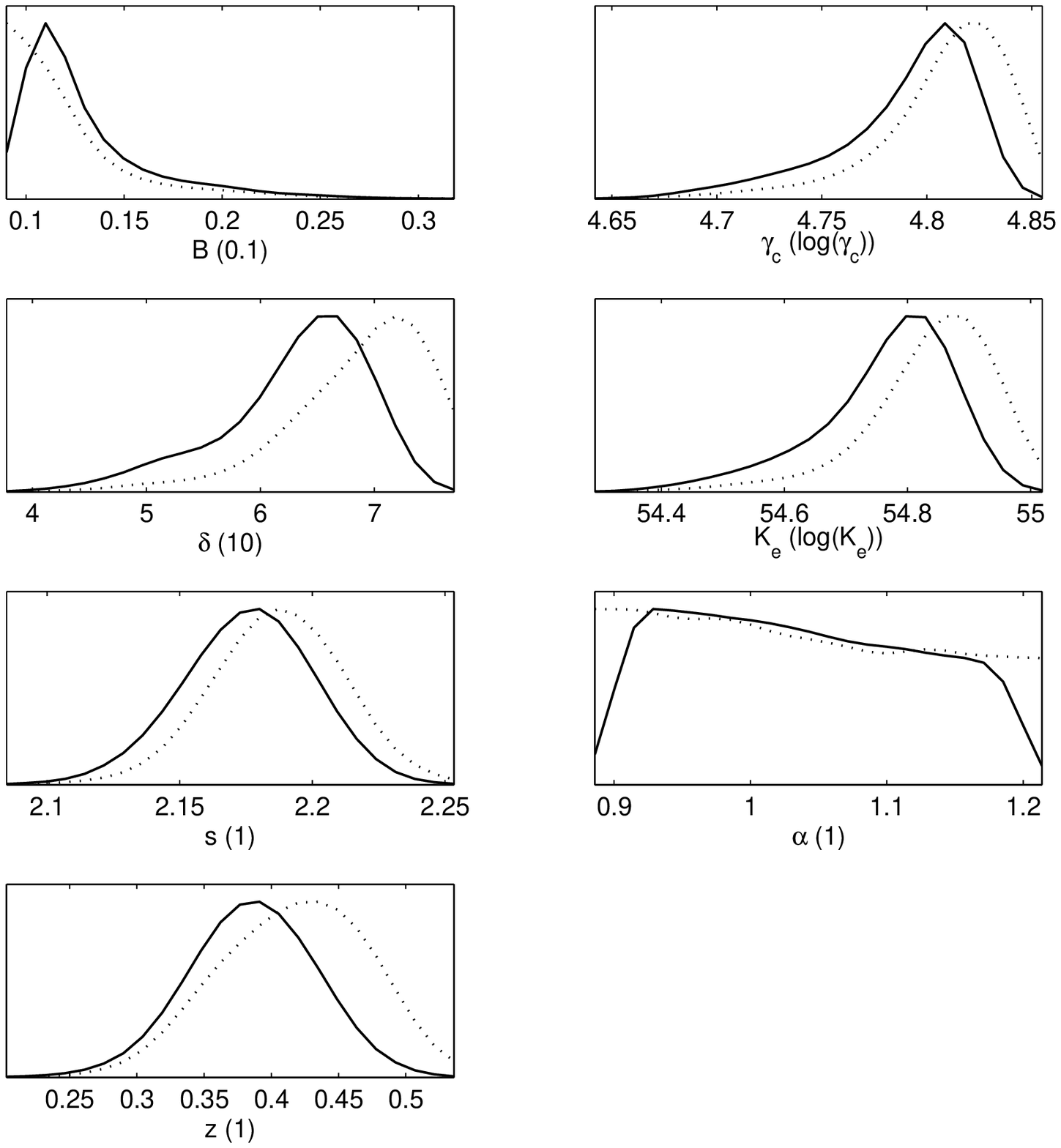}
\end{minipage}%
}
\hspace{1cm}
\subfigure[The correlations of the parameters in PLC EEDs]{
 \begin{minipage}[t]{8cm}
 \centering
\includegraphics[height=7cm,width=8cm]{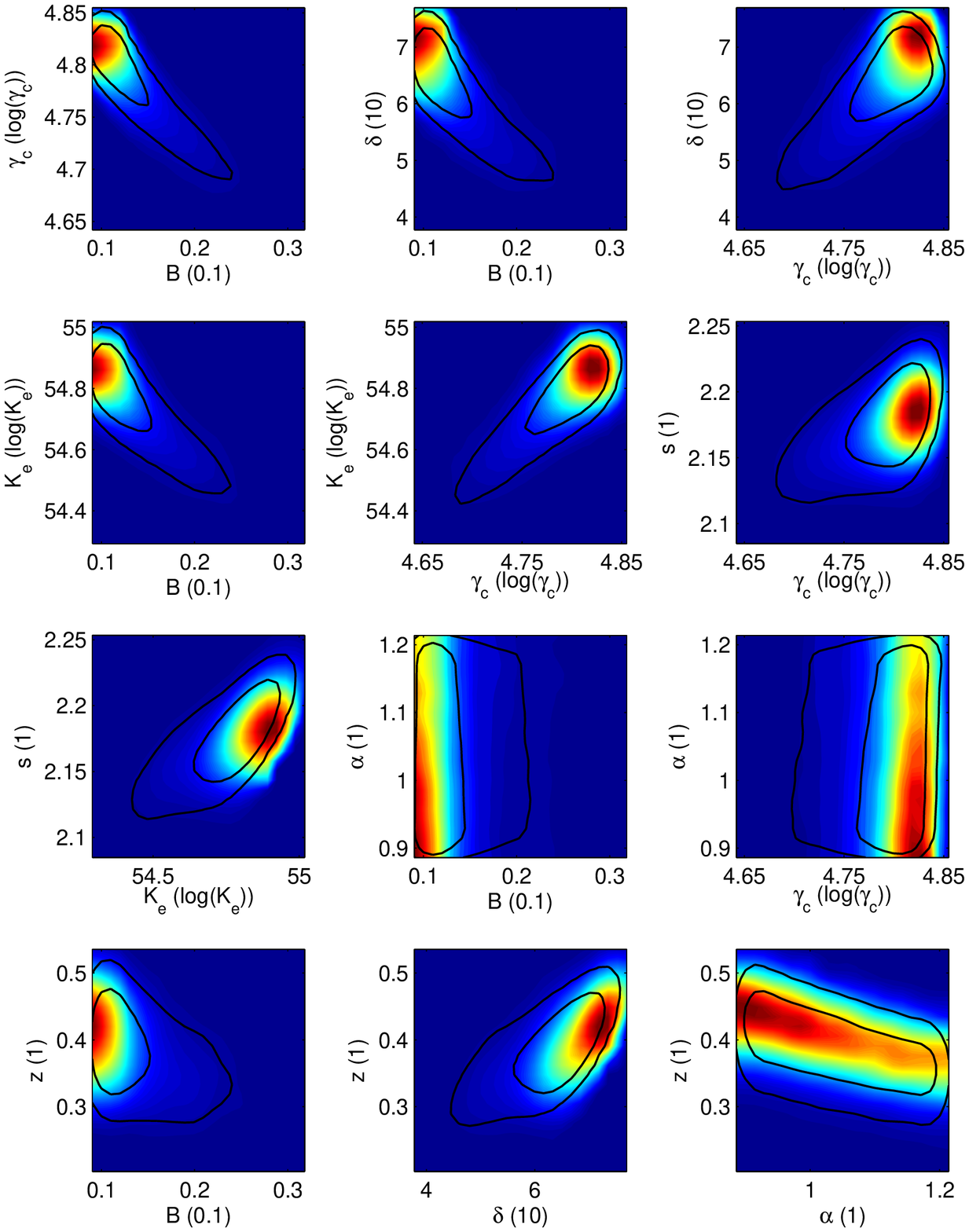}%
 \centering
\end{minipage}%
}%

\caption{3C66A. Left panels: the distributions of the parameters given by three types of EEDs, where the dotted lines show the maximum likelihood distributions, the solid lines show the marginalized probability distributions. Right panels: 2-D confident contours of the parameter correlations for three types of EEDs, where the contours of the parameters with larger correlations are shown. The inner contours denote the 68\% C.L., while the outer contours represent the 95\% C.L.}
\end{figure*}

\begin{figure*}
\subfigure[The distributions of the parameters in BPL EEDs]{
 \begin{minipage}[t]{8cm}
 \centering
\includegraphics[height=7cm,width=8.0cm]{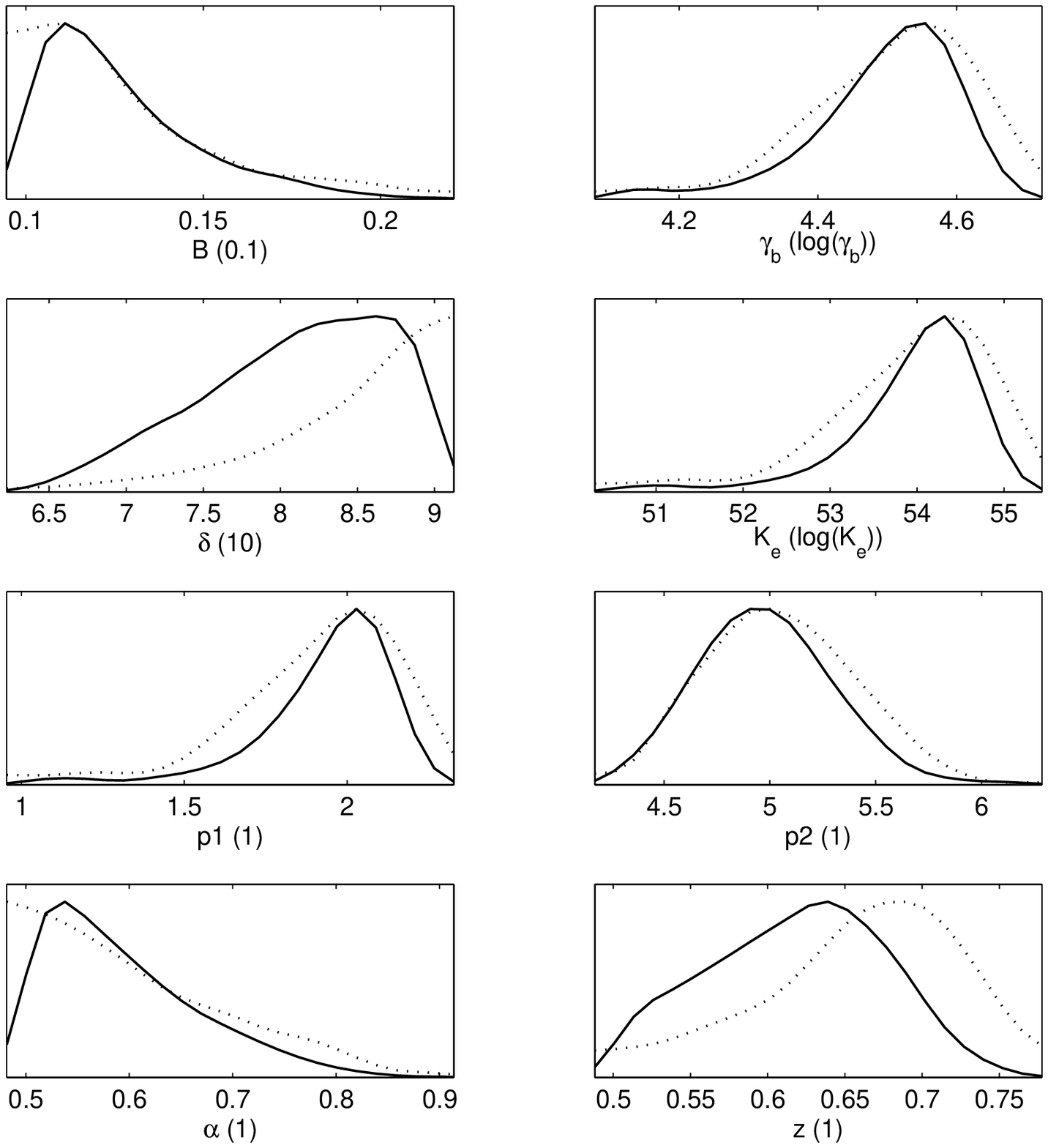}
\end{minipage}%
}
\hspace{1cm}
\subfigure[The correlations of the parameters in BPL EEDs]{
 \begin{minipage}[t]{8cm}
 \centering
\includegraphics[height=7cm,width=8.0cm]{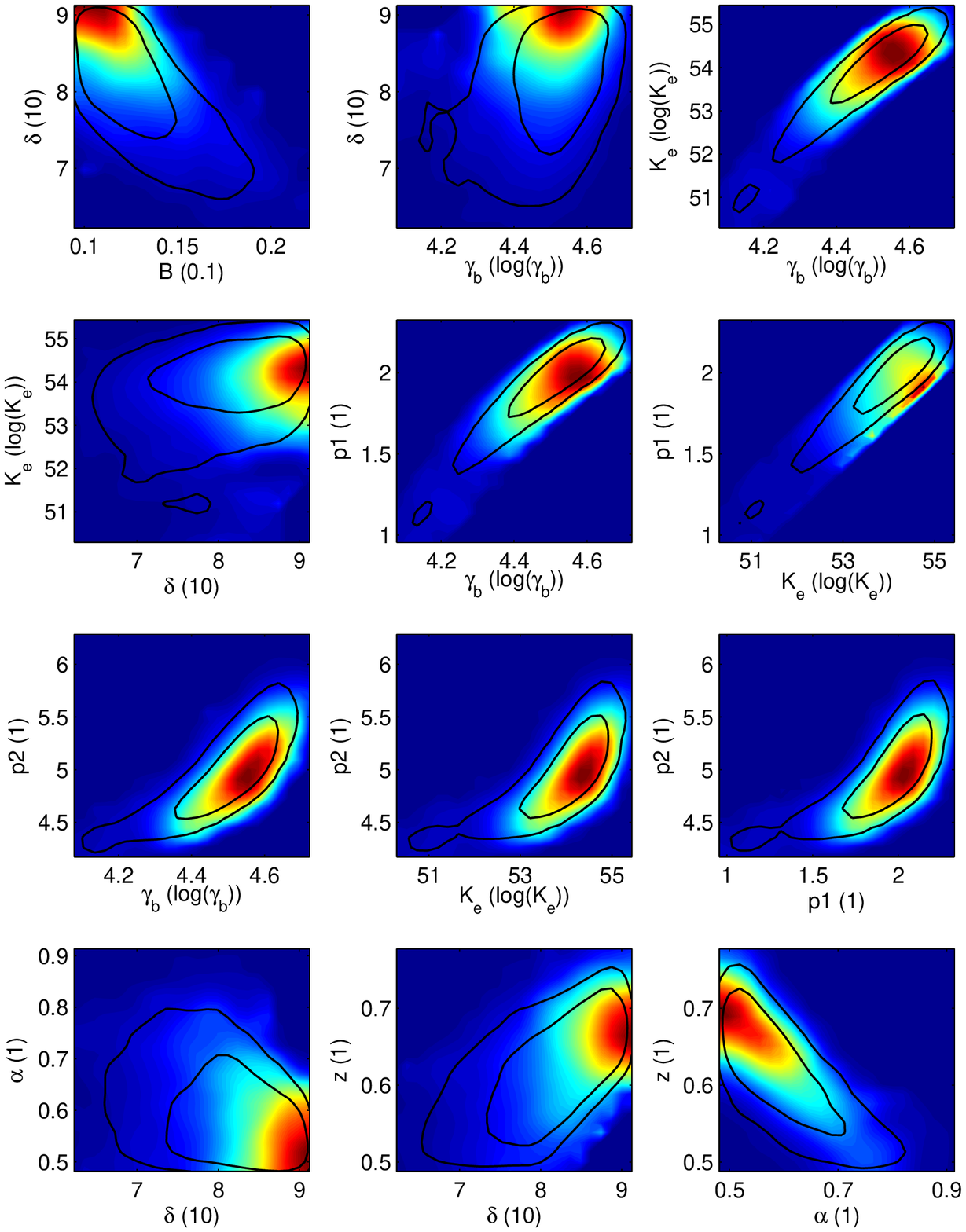}
\end{minipage}%
}
\hspace{1cm}
\subfigure[The distributions of the parameters in PLLP EEDs]{
 \begin{minipage}[t]{8cm}
 \centering
\includegraphics[height=7cm,width=8cm]{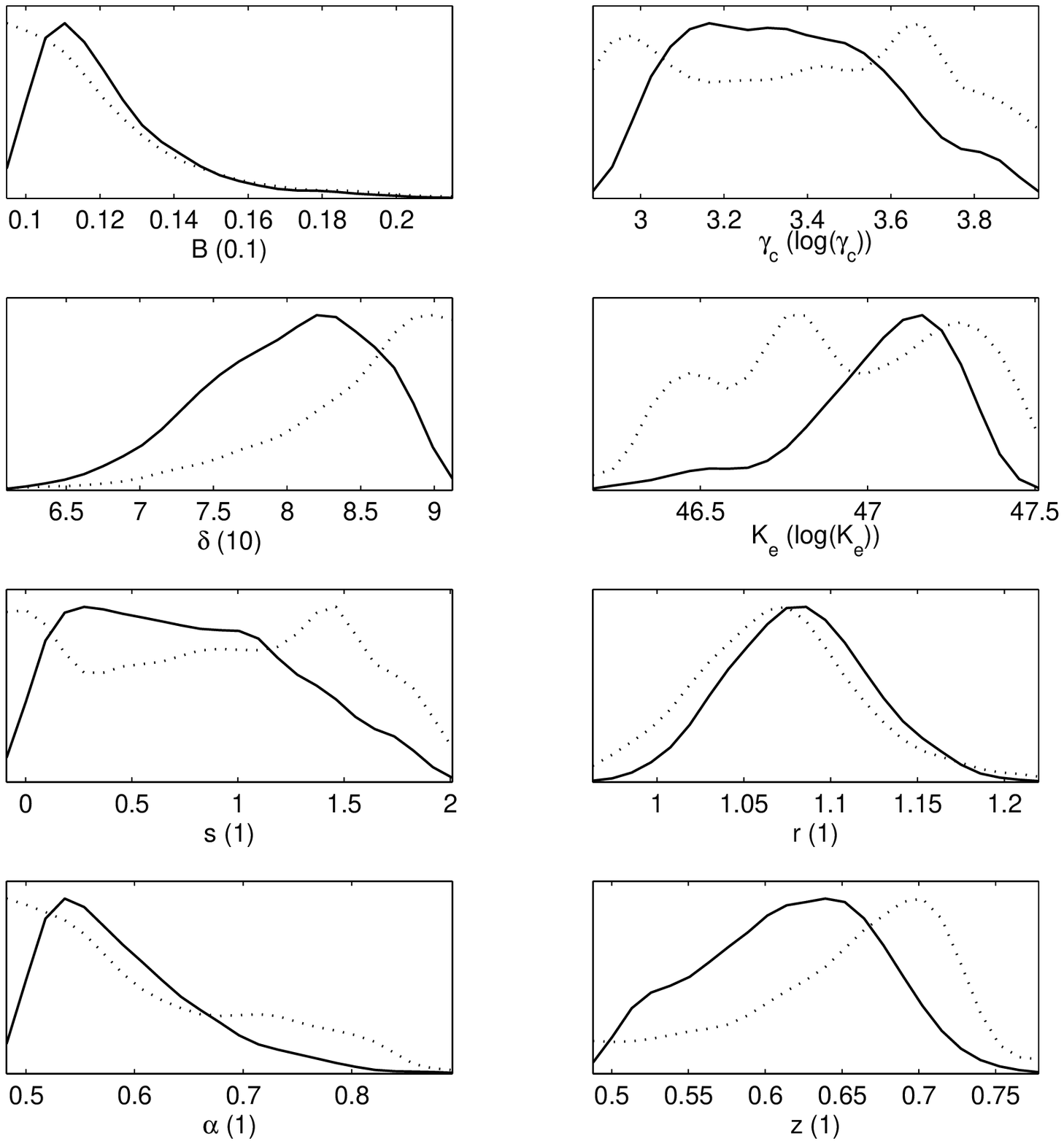}
\end{minipage}%
}
\hspace{1cm}
\subfigure[The correlations of the parameters in PLLP EEDs]{
 \begin{minipage}[t]{8cm}
 \centering
\includegraphics[height=7cm,width=8cm]{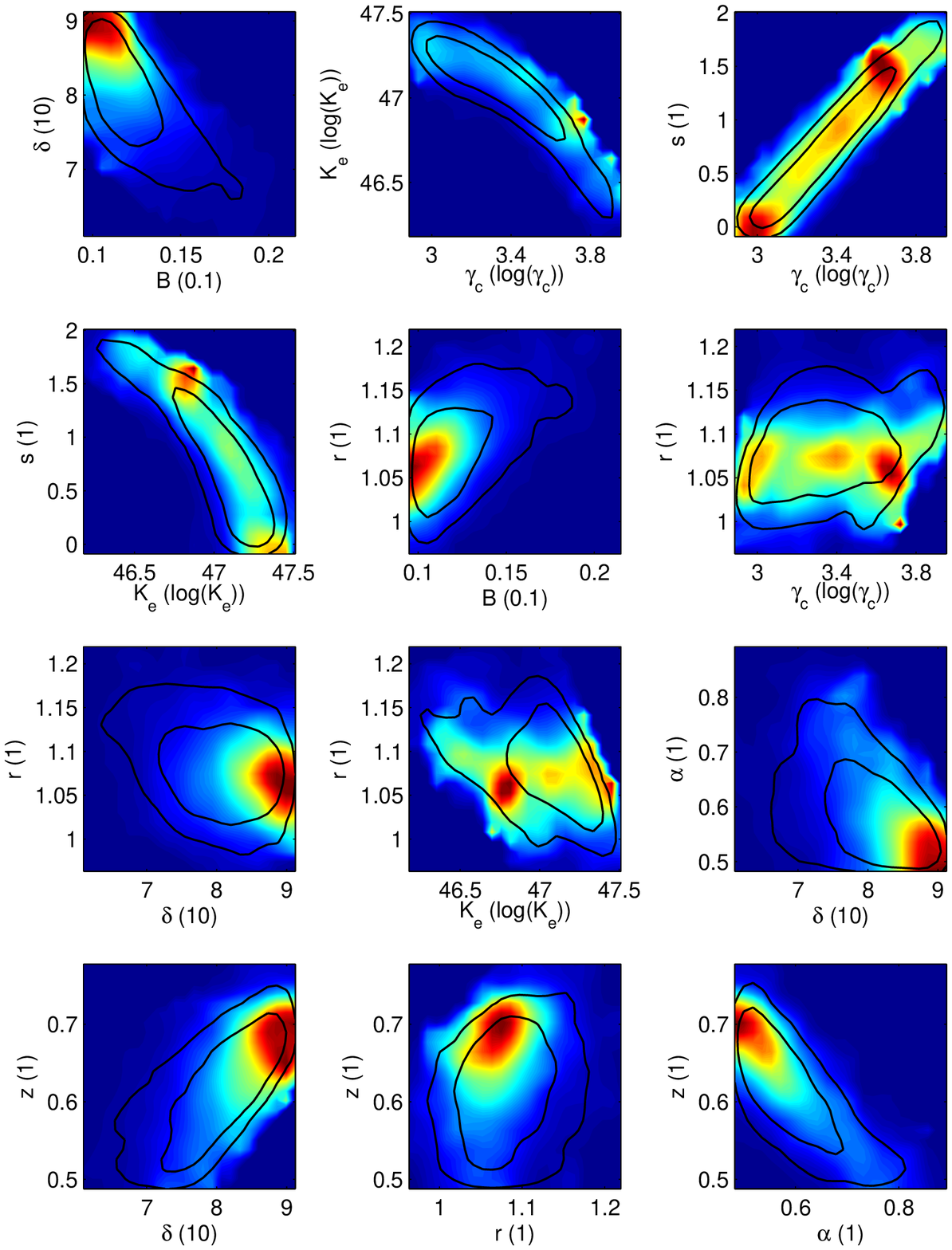}
\end{minipage}%
}
\hspace{1cm}
\subfigure[The distributions of the parameters in PLC EEDs]{
 \begin{minipage}[t]{8cm}
 \centering
\includegraphics[height=7cm,width=8cm]{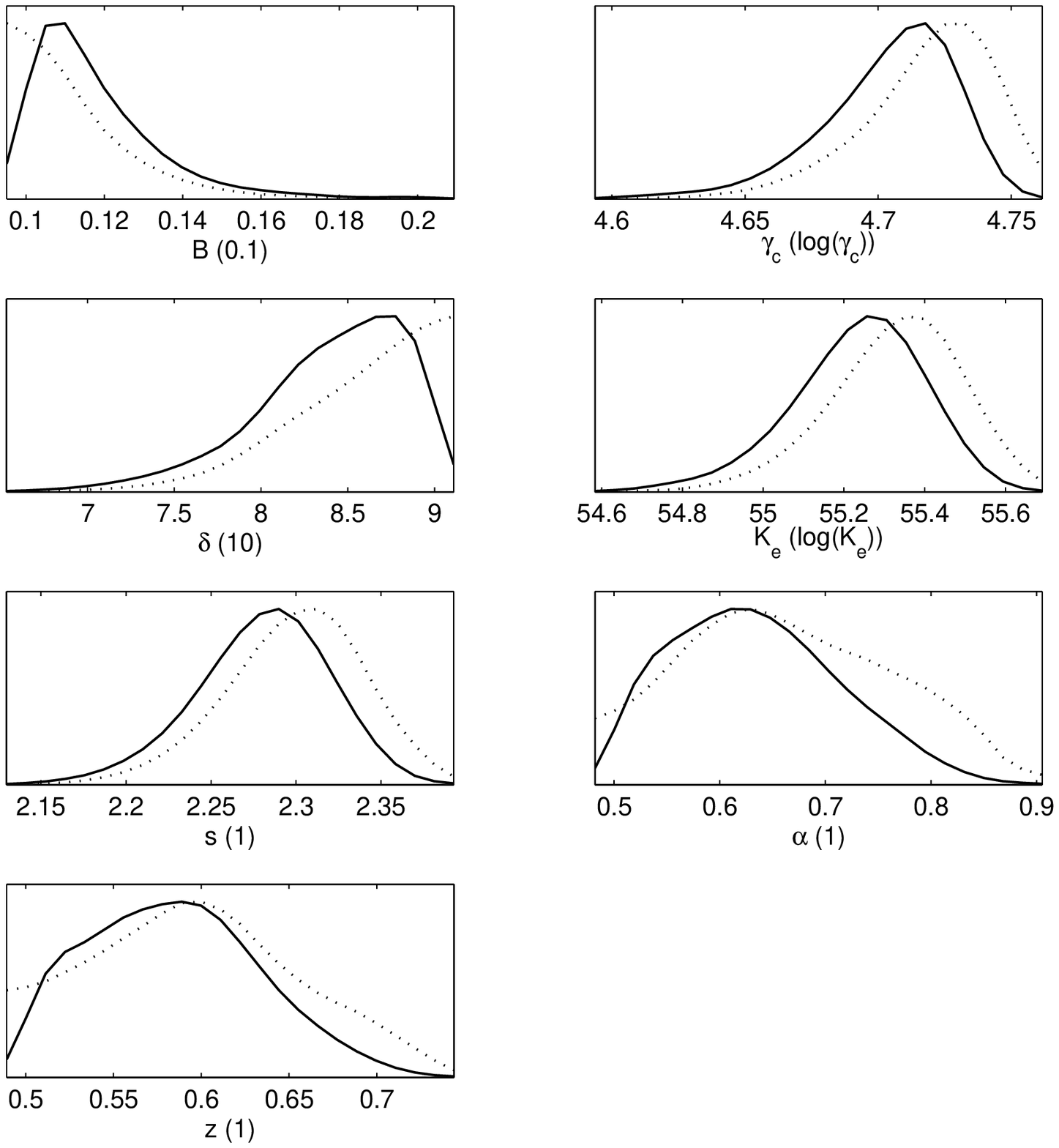}
\end{minipage}%
}
\hspace{1cm}
\subfigure[The correlations of the parameters in PLC EEDs]{
 \begin{minipage}[t]{8cm}
 \centering
\includegraphics[height=7cm,width=8cm]{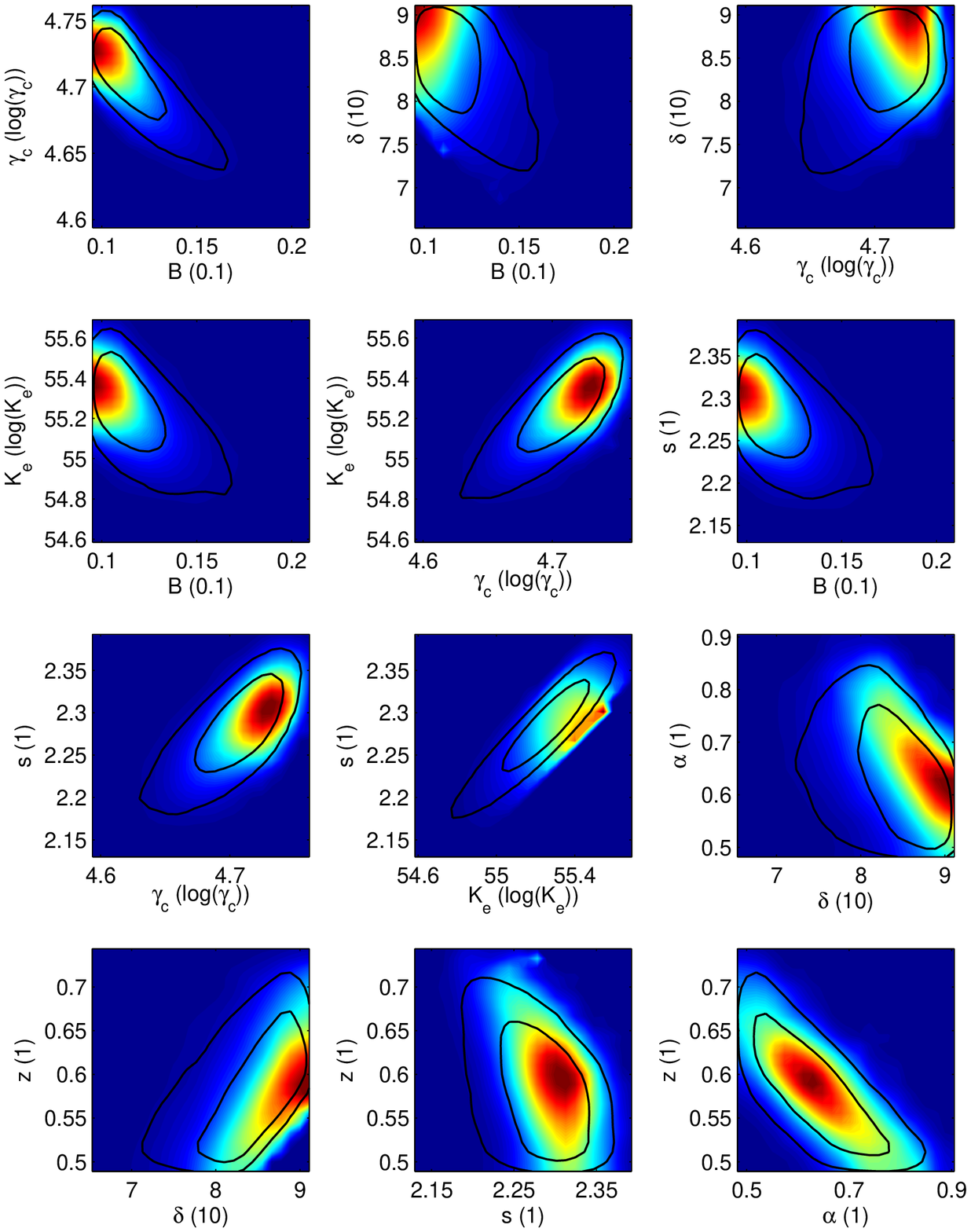}%
 \centering
\end{minipage}%
}%

\center{\textbf{Figure 2 --Continue}. PKS1424+240. Same as in Fig.2.}
\end{figure*}

\begin{figure*}
\subfigure[The distributions of the parameters in BPL EEDs]{
 \begin{minipage}[t]{8cm}
 \centering
\includegraphics[height=7cm,width=8.0cm]{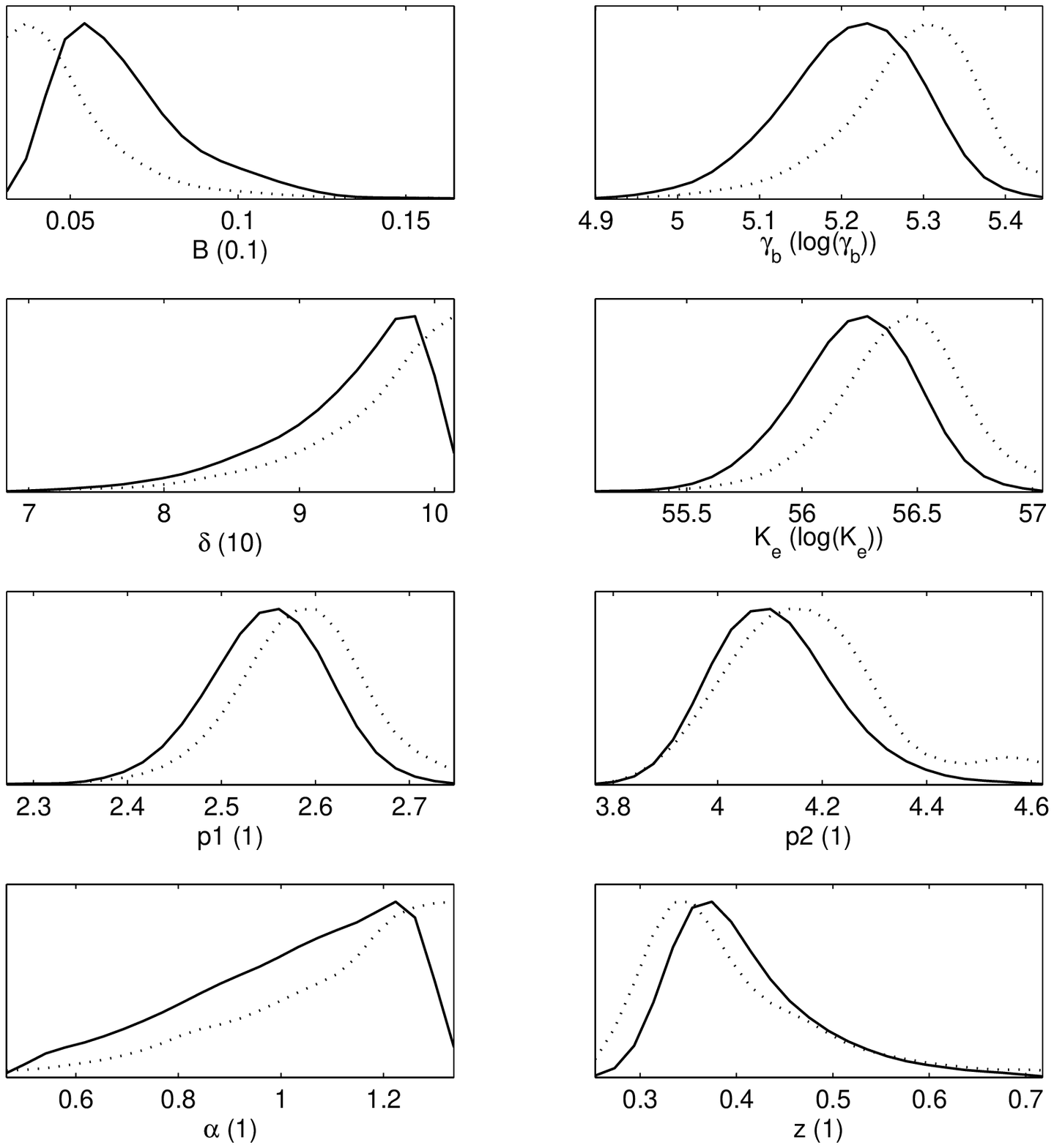}
\end{minipage}%
}
\hspace{1cm}
\subfigure[The correlations of the parameters in BPL EEDs]{
 \begin{minipage}[t]{8cm}
 \centering
\includegraphics[height=7cm,width=8.0cm]{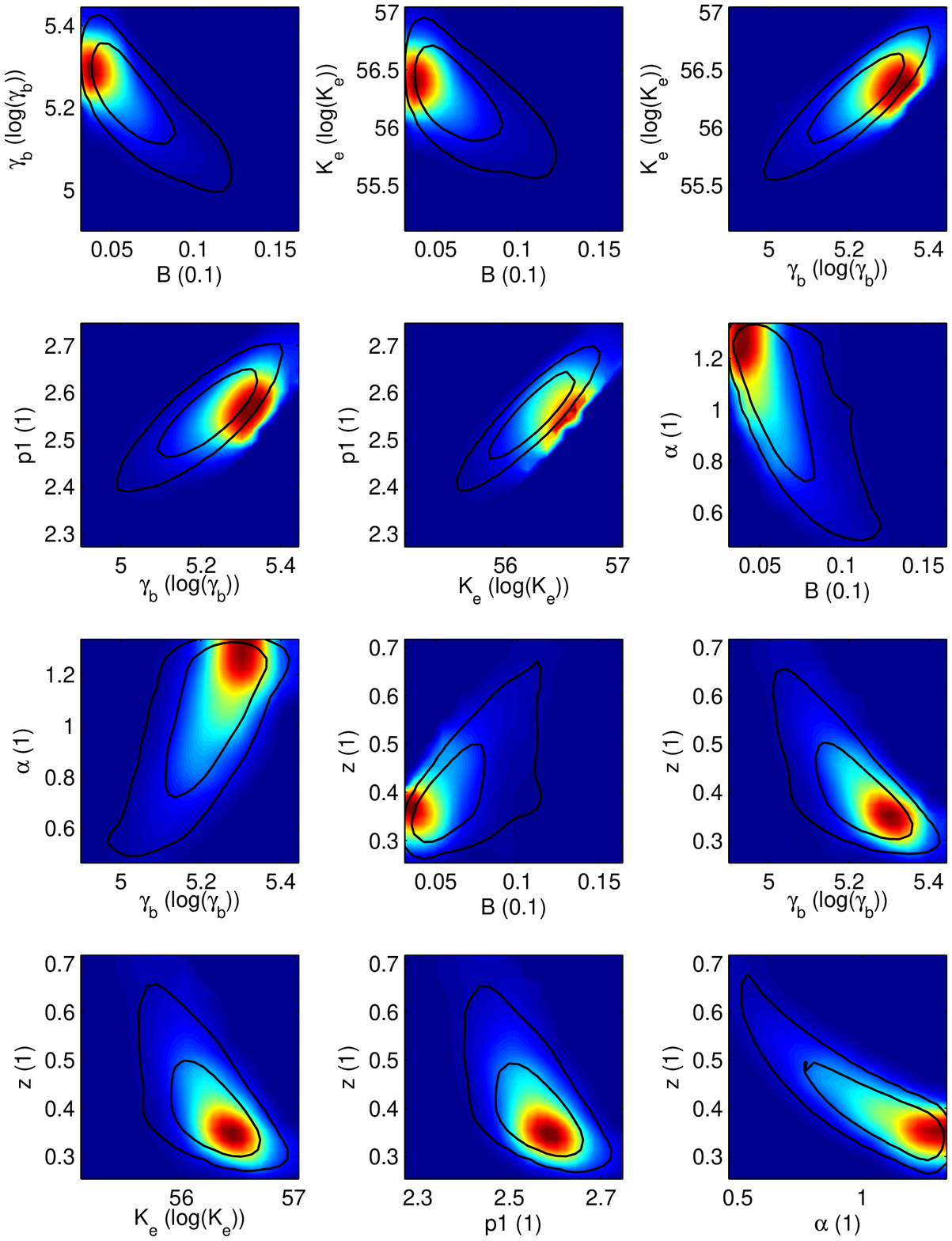}
\end{minipage}%
}
\hspace{1cm}
\subfigure[The distributions of the parameters in PLLP EEDs]{
 \begin{minipage}[t]{8cm}
 \centering
\includegraphics[height=7cm,width=8cm]{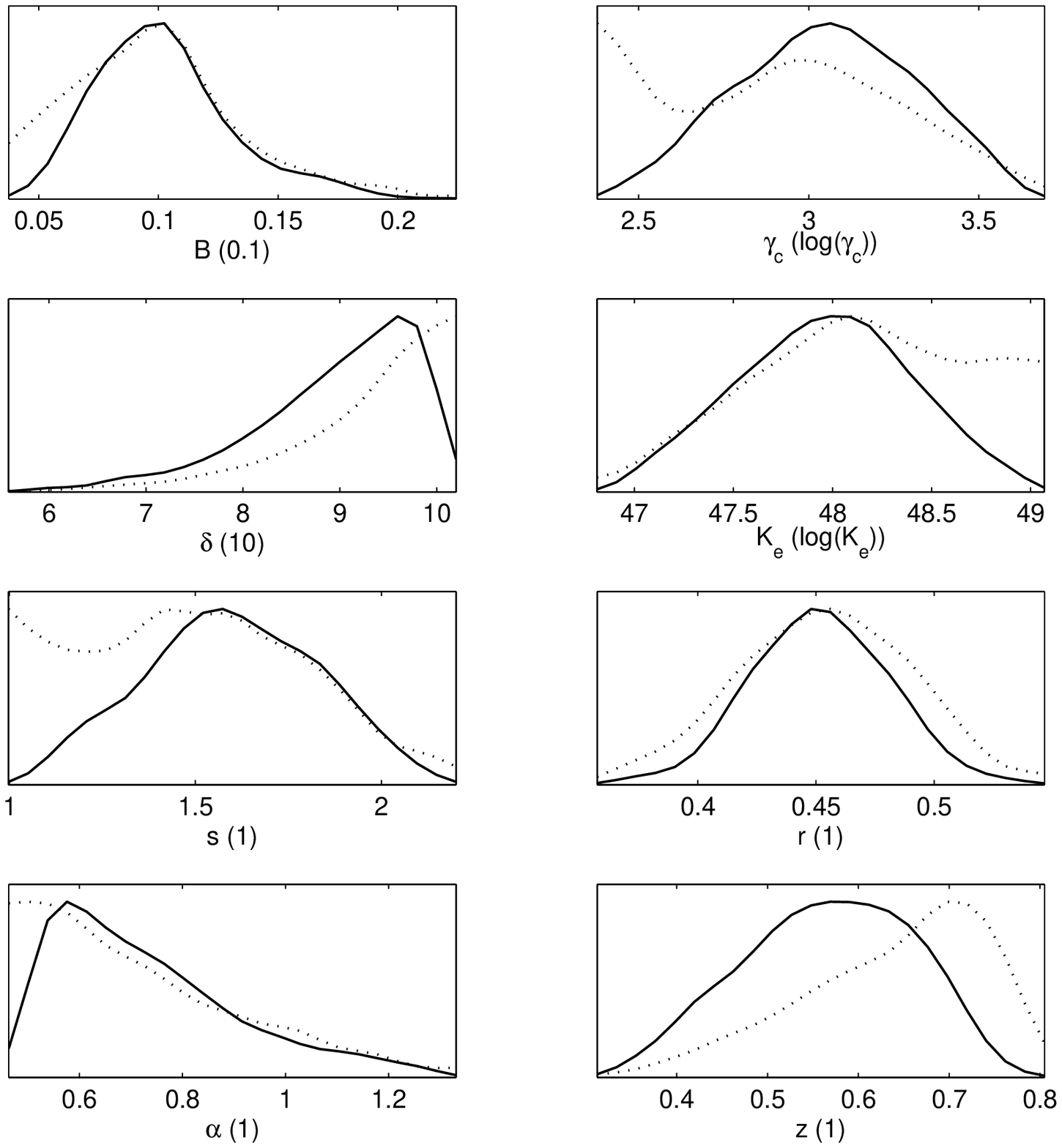}
\end{minipage}%
}
\hspace{1cm}
\subfigure[The correlations of the parameters in PLLP EEDs]{
 \begin{minipage}[t]{8cm}
 \centering
\includegraphics[height=7cm,width=8cm]{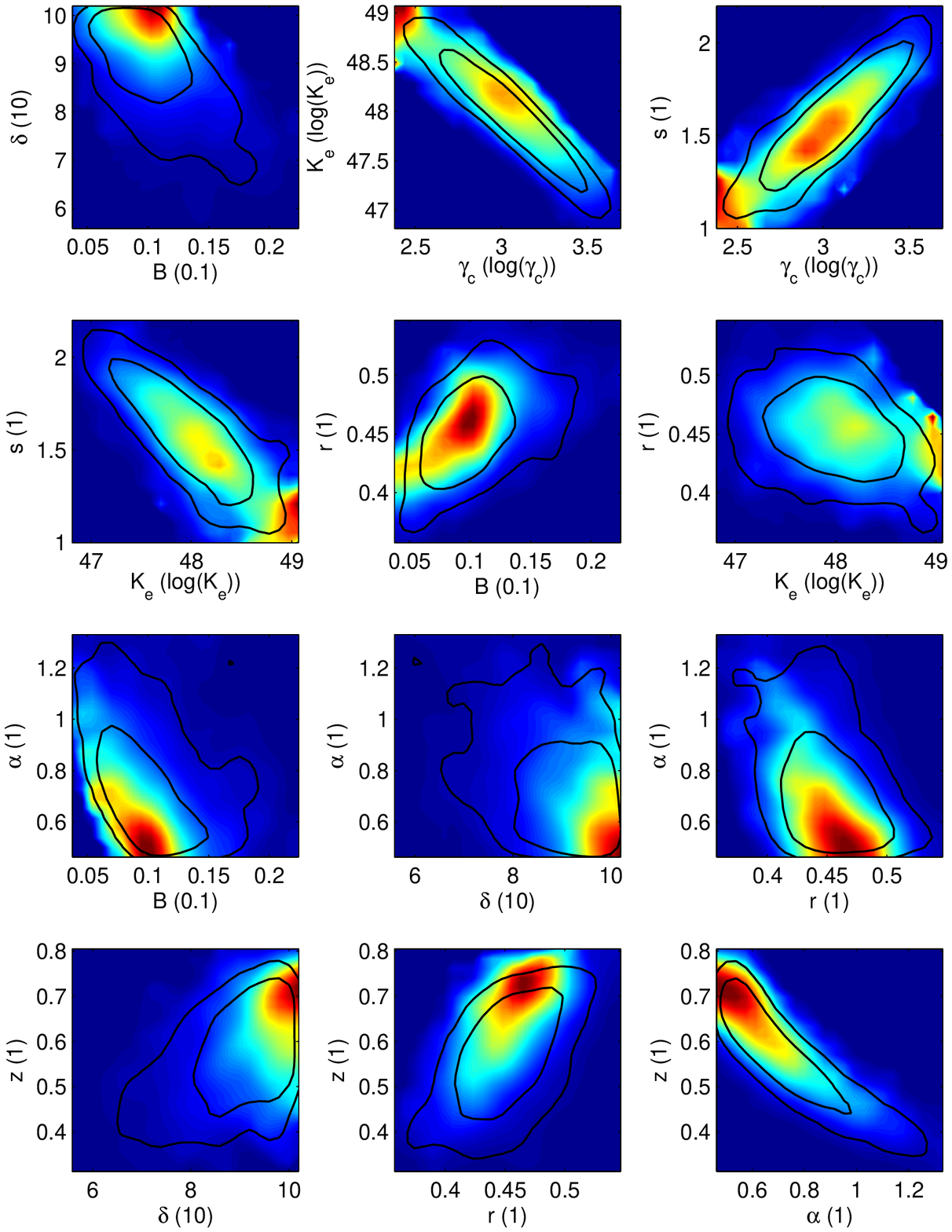}
\end{minipage}%
}
\hspace{1cm}
\subfigure[The distributions of the parameters in PLC EEDs]{
 \begin{minipage}[t]{8cm}
 \centering
\includegraphics[height=7cm,width=8cm]{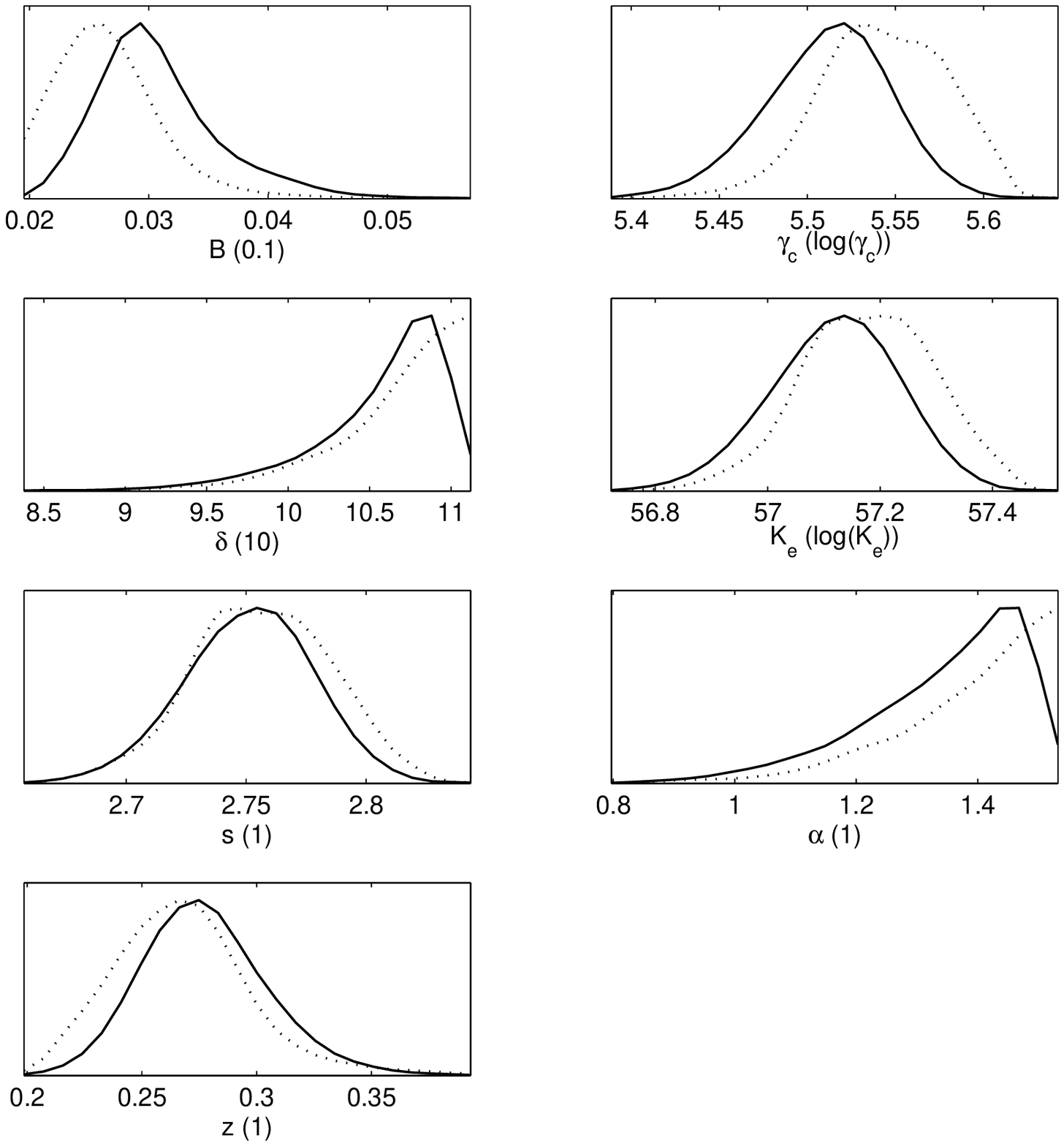}
\end{minipage}%
}
\hspace{1cm}
\subfigure[The correlations of the parameters in PLC EEDs]{
 \begin{minipage}[t]{8cm}
 \centering
\includegraphics[height=7cm,width=8cm]{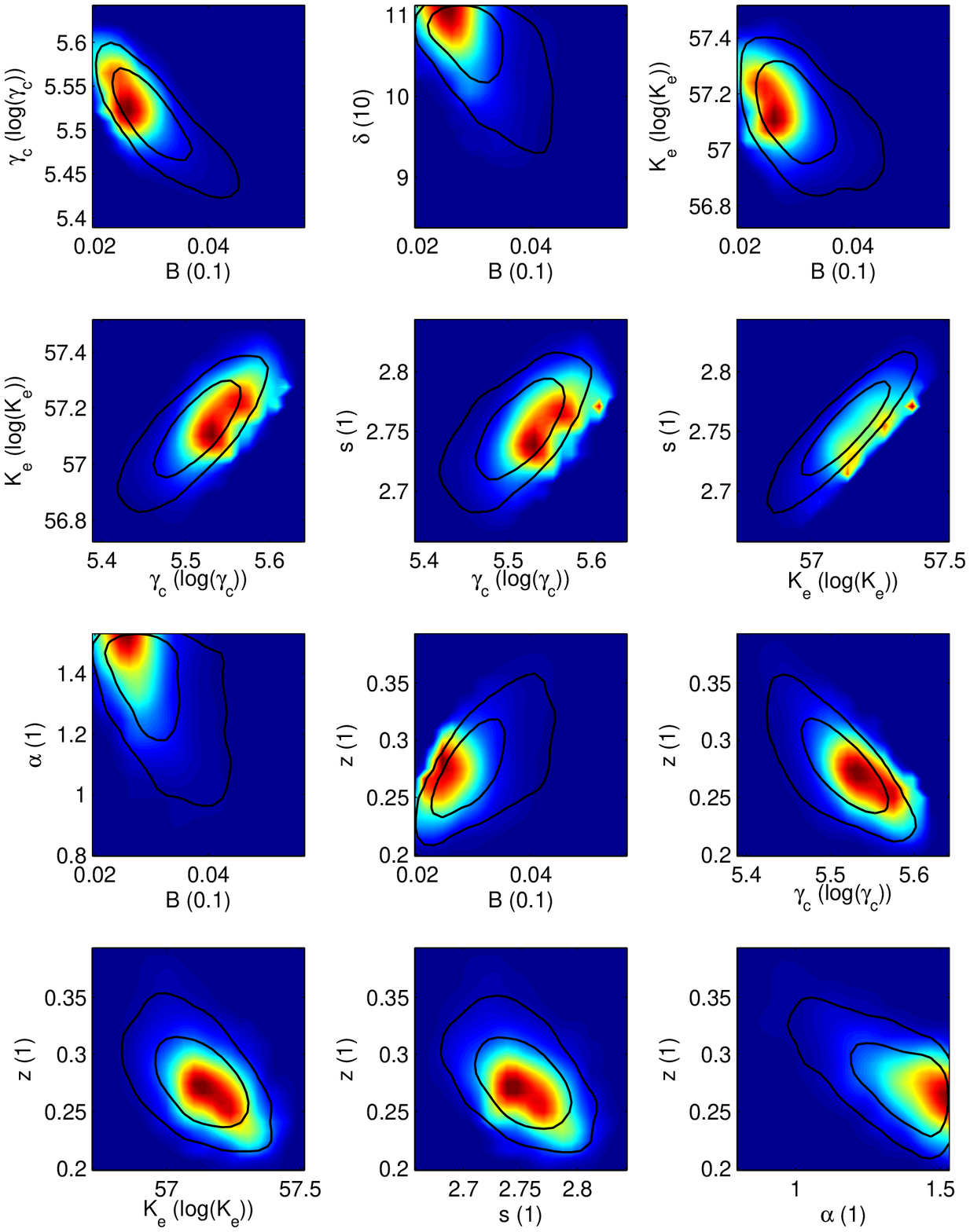}%
 \centering
\end{minipage}%
}%

\center{\textbf{Figure 2 --Continue}. PKS1424+240 flare. Same as in Fig.2.}
\end{figure*}

\begin{figure*}
\subfigure[The distributions of the parameters in BPL EEDs]{
 \begin{minipage}[t]{8cm}
 \centering
\includegraphics[height=7cm,width=8.0cm]{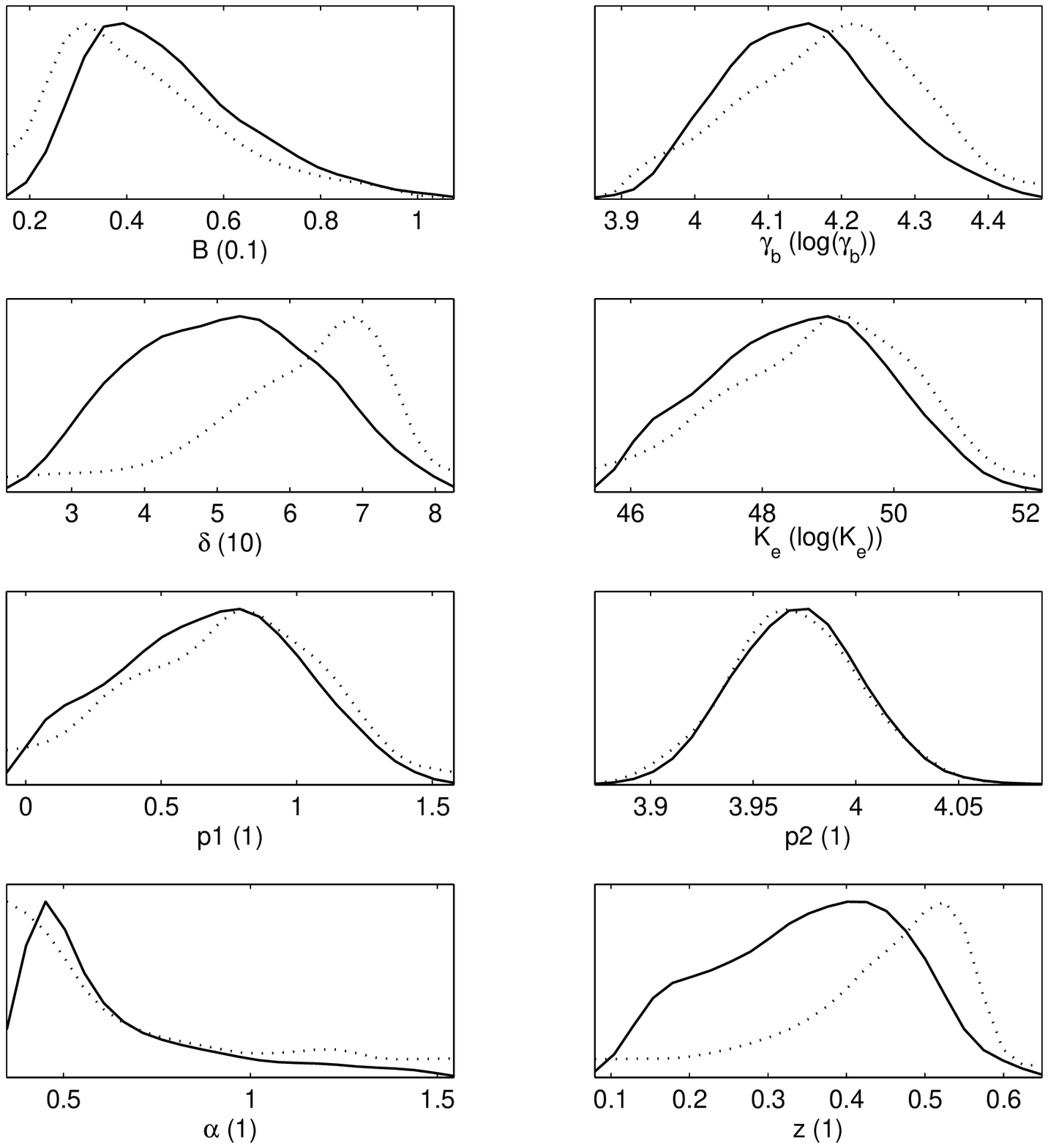}
\end{minipage}%
}
\hspace{1cm}
\subfigure[The correlations of the parameters in BPL EEDs]{
 \begin{minipage}[t]{8cm}
 \centering
\includegraphics[height=7cm,width=8.0cm]{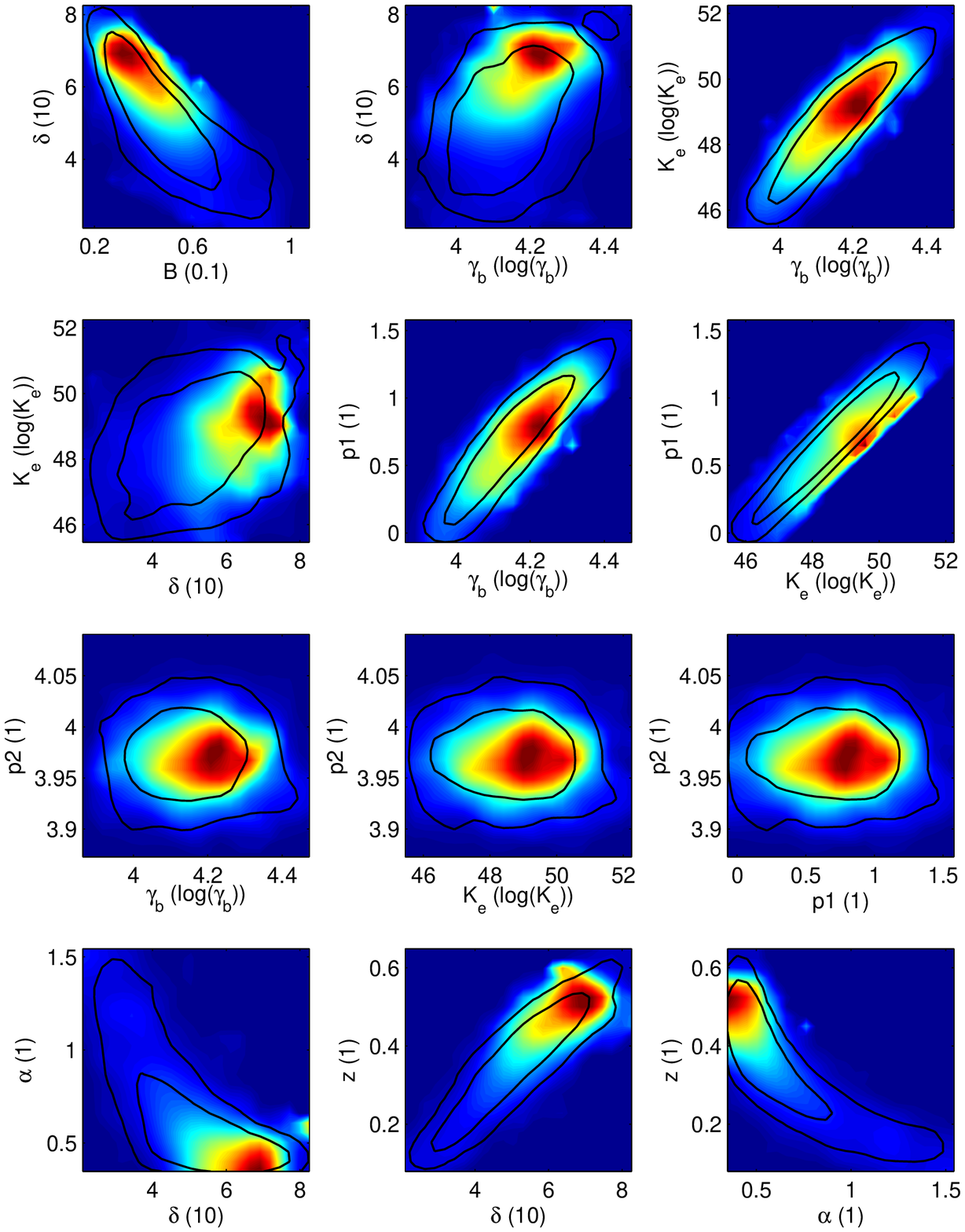}
\end{minipage}%
}
\hspace{1cm}
\subfigure[The distributions of the parameters in PLLP EEDs]{
 \begin{minipage}[t]{8cm}
 \centering
\includegraphics[height=7cm,width=8cm]{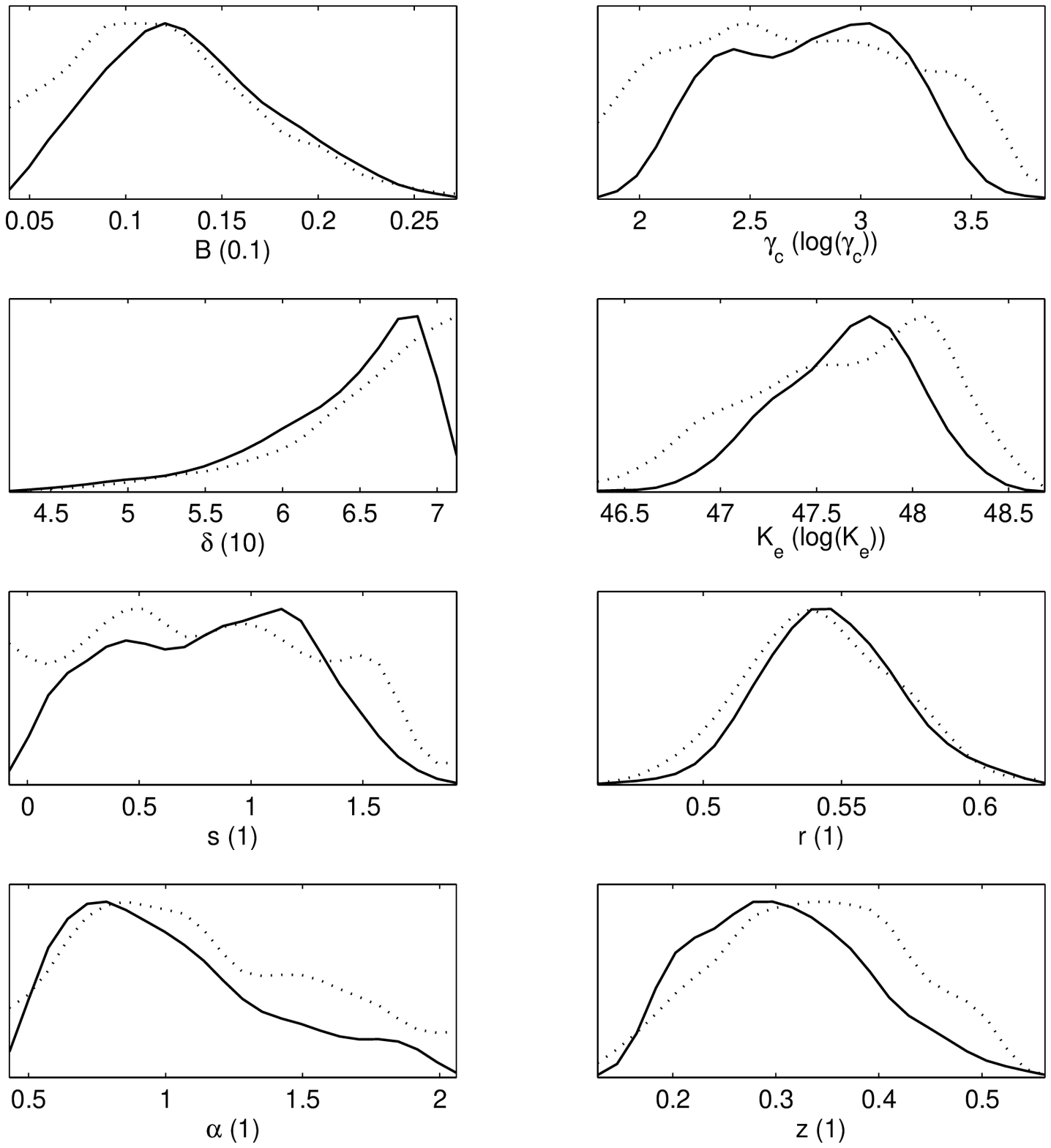}
\end{minipage}%
}
\hspace{1cm}
\subfigure[The correlations of the parameters in PLLP EEDs]{
 \begin{minipage}[t]{8cm}
 \centering
\includegraphics[height=7cm,width=8cm]{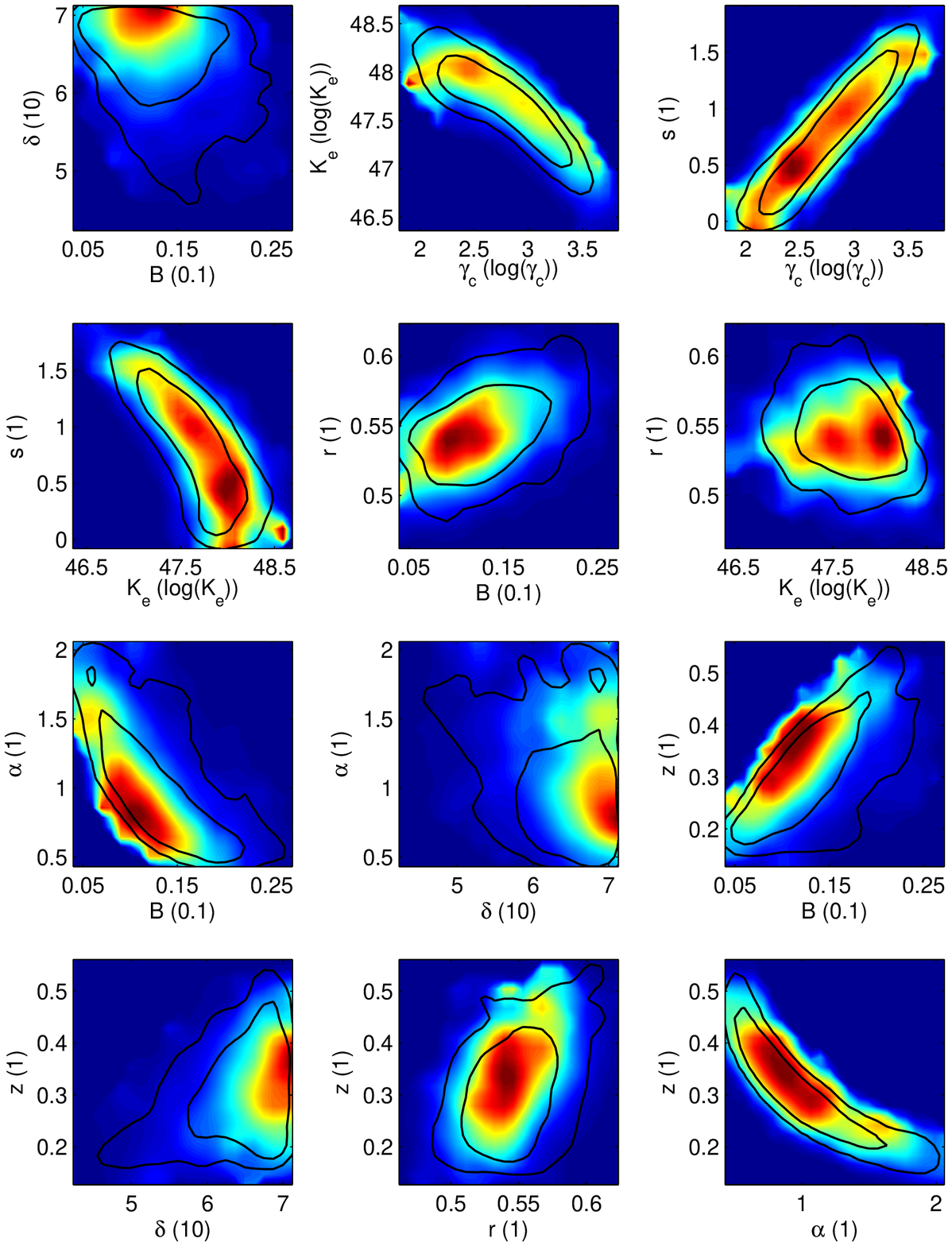}
\end{minipage}%
}
\hspace{1cm}
\subfigure[The distributions of the parameters in PLC EEDs]{
 \begin{minipage}[t]{8cm}
 \centering
\includegraphics[height=7cm,width=8cm]{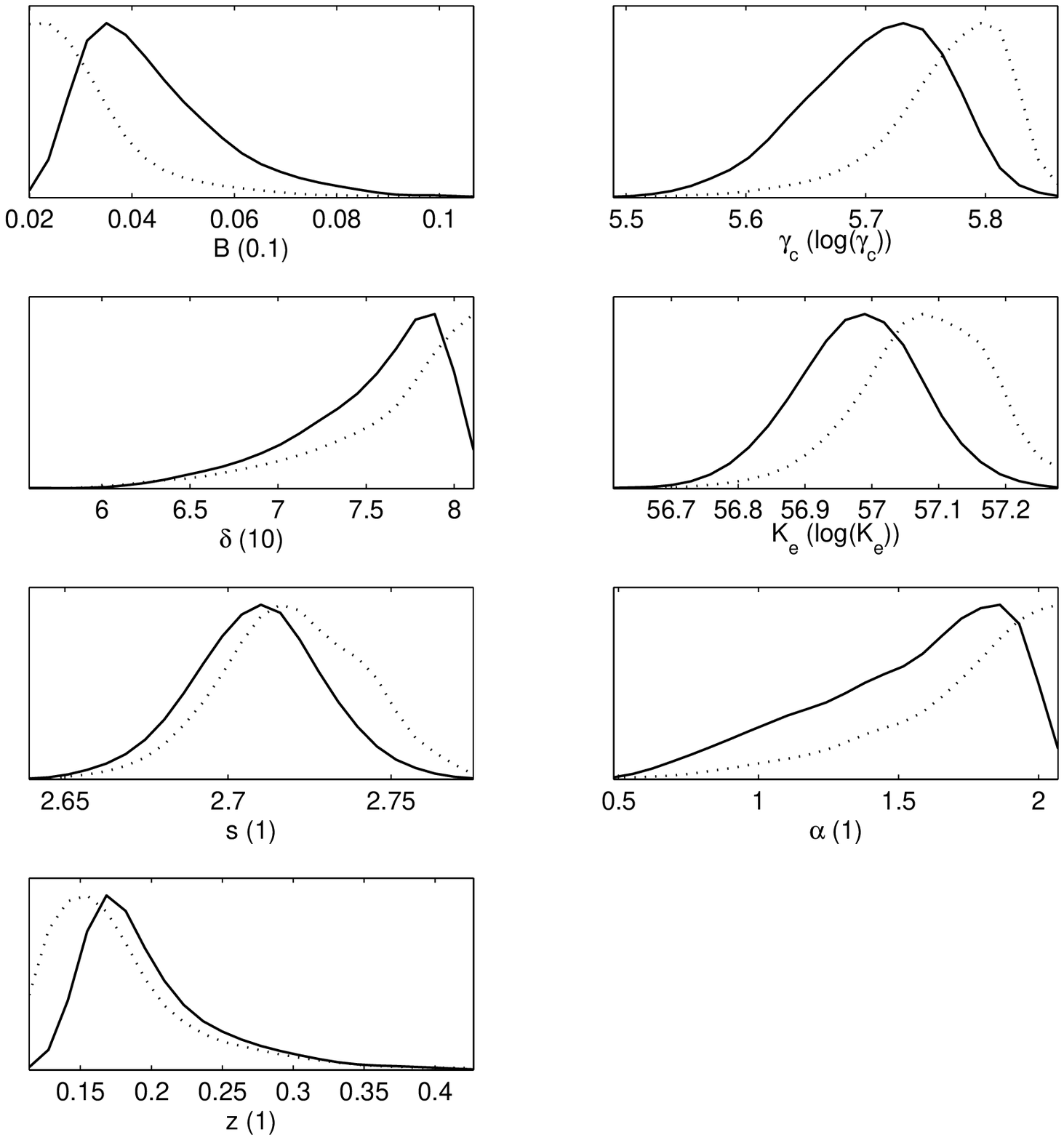}
\end{minipage}%
}
\hspace{1cm}
\subfigure[The correlations of the parameters in PLC EEDs]{
 \begin{minipage}[t]{8cm}
 \centering
\includegraphics[height=7cm,width=8cm]{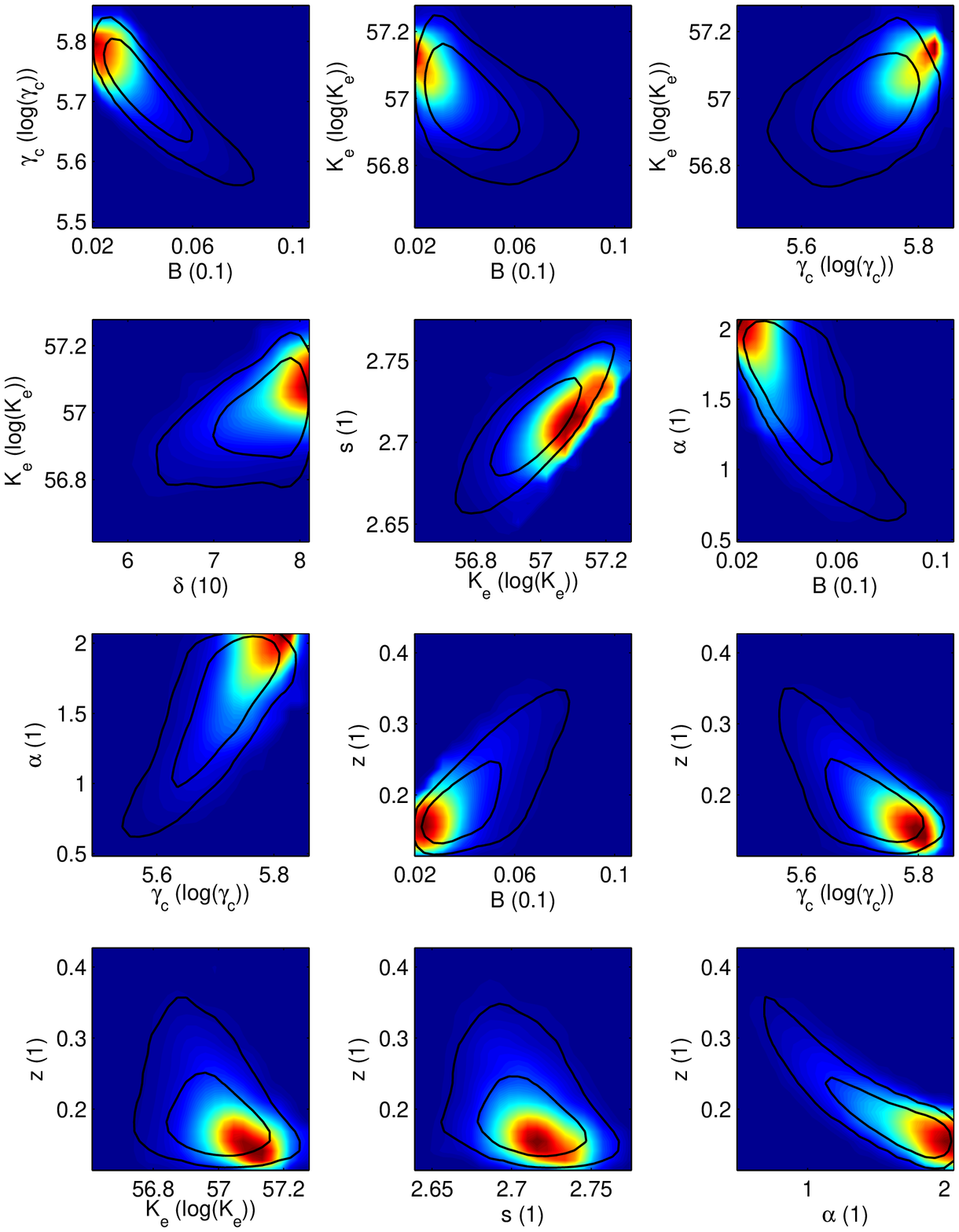}%
 \centering
\end{minipage}%
}%

\center{\textbf{Figure 2 --Continue}. PG1553+113. Same as in Fig.2.}
\end{figure*}

\section{DISCUSSION AND CONCLUSION}
In the paper, we use the MCMC method and the EBL model with a normalization parameter to estimate the redshifts of TeV BL Lacs by fitting the quasi-simultaneous SEDs based on three types of EEDs. We also give the redshifts of TeV BL Lac in the marginalized 68\% C.L.

From the Fig.1-2, the SEDs fitted by the PLC EED do not cover the observed SEDs very well, in which the maximum likelihood distributions of the parameters are not consistent with the marginalized distributions, showing that the PLC EED is not good to model the SEDs. This result implies that shock acceleration could not work or stochastic acceleration should dominate over radiative cooling in the emitting blob. From the 2D marginalized probability distributions in Fig.2, we find that $B$ and $\delta$ are anti-correlation. According to the SSC model, $B\delta \propto [\nu_{\text{sy}}^{2}/\nu_{\text{IC}}](1+z)$ in the Thomson regime and $B/\delta \propto [\nu_{\text{sy}}/\nu_{\text{IC}}^{2}]/(1+z)$ in the Klein-Nishina (KN) regime \citep{tav1998}. Our result is in agreement with the SSC model prediction and suggests that IC occurs in the Thomson regime.  From the table 1-3, we find that a redshift dependence of the EBL
normalization factor $\alpha$ appears, where $\alpha > 1$ for lower redshift sources and $\alpha < 1$ for higher redshift sources. This property is also supported by the 2D distributions in Fig.2, in which $\alpha$ and redshift are anti-correlation in three types of EEDs except for 3C66A in the BPL EEDs.

It is noted that our method relies on the assumption: (1) both HE and VHE $\gamma$ -rays are produced by SSC process in same region; (2) the (quasi-) simultaneous multi-waveband SEDs are needed. In the paper, the SEDs of three TeV BL Lacs are fitted by three types of EEDs in one-zone SSC model. We find that the fitting SEDs shows a significant distinction in 20 KeV- 50 KeV bands for three types of EEDs, implying that hard X-rays can be used to distinguish EEDs. Because the one-zone SSC is well used in the GeV-TeV BL Lacs \citep{Ghi2010b,zhang2012}, in which three types of EEDs are commonly used to fit the SEDs of blazars, our method is a useful way to constrain the redshifts of blazars when the (quasi-) simultaneous multi-waveband data are well acquired.

\section*{Acknowledgments}
We thank anonymous referee for useful comments and suggestions. This research has used the NASA/IPAC Extragalactic Database (NED) which is operated by the Jet Propulsion Laboratory,
California Institute of Technology, under contract with National Aeronautics and Space Administration.
The authors gratefully acknowledge the financial supports from the
National Natural Science Foundation of China 11673060, 11661161010,11603066, and the Natural Science Foundation
of Yunnan Province under grant 2016FB003. The authors gratefully acknowledge the computing time granted by the Yunnan Observatories,
and provided on the facilities at the Yunnan
Observatories Supercomputing Platform. Zhunli Yuan and Dahai Yan are supported by the CAS ``Light of West China" Program.




\clearpage

\end{document}